\documentclass[12pt]{article}
\usepackage{epsfig, amssymb}
\usepackage{graphicx,epsfig}
\usepackage{color}
\begin{document}
\begin{titlepage}
\thispagestyle{empty}
\begin{flushright}
\end{flushright}

\bigskip

\begin{center}
\noindent{\Large \textbf
{4-point function from conformally coupled scalar in AdS$_6$}}\\
\vspace{2cm} \noindent{
Jae-Hyuk Oh${}^{a}$\footnote{jack.jaehyuk.oh@gmail.com, jaehyukoh@hanyang.ac.kr} 
}

\vspace{1cm}
  {\it
Department of Physics, Hanyang University \\
 Seoul 133-891, Korea${}^{a}$\\
 }
\end{center}

\vspace{0.3cm}
\begin{abstract}
We explore conformally coupled scalar theory in AdS$_{6}$ extensively and their classical solutions
by employing power expansion order by order in
its self-interaction coupling $\lambda$. We describe how we get the classical solutions by diagrammatic ways which show general rules constructing the classical solutions.
We study holographic correlation functions of scalar operator deformations to a certain 5-dimensional conformal field theory
where the operators share the same scaling dimension $\Delta=3$, 
from the classical solutions. 
We do not assume any specific form of the micro Lagrangian density of the 5-dimensional conformal field theory.
For our solutions, we choose a scheme where we remove co-linear divergences of momenta along the AdS boundary directions which frequently appear in the classical solutions. This shows clearly that the holographic correlation functions are free from the co-linear divergences.
It turns out that this theory provides correct conformal 2- and 3- point functions of the $\Delta=3$ scalar operators as expected in previous literature. It makes sense since 2- and 3- point functions are determined by global conformal symmetry not being dependent on the details of the conformal theory.
We also get 4-point function from this holographic model. 
In fact, it turns out that the 4-point correlation function is not conformal because it does not satisfy the special conformal Ward identity although it does dilation Ward identity and respect $SO(5)$ rotation symmetry. However, in the co-linear limit that all the external momenta are in a same direction, the 4-point function is conformal which means that it satisfy the special conformal Ward identity. 
We inspect holographic $n$-point functions of this theory which can be obtained by employing a certain Feynman-like rule. This rule is a 
construction of $n$-point function by connecting $l$-point functions each other where $l<n$. In the co-linear limit, these $n$-point functions reproduce the conformal $n$-point functions of $\Delta=3$ scalar operators in $d=5$ Euclidean space addressed in arXiv:2001.05379.

\end{abstract}
\end{titlepage}

\newpage

\tableofcontents
\section{Introduction}
AdS/CFT correspondence has shed light on strongly coupled field theories such as quark gluon plasma, conformal fluid dynamics, condensed matter theories and so on. Together with these issues, holography itself is suggested as a wide field to study on, especially to look at its UV-IR structure(relation), by employing for example, sliding membrane, holographic renormalization group. One of the interesting branches of studies on holography is to develop holographic conformal $n$-point functions in momentum space. To get 
conformal $n$-point functions (in momentum space), one needs to solve conformal Ward identities, among them the special conformal Ward identity is a second order differential equation with respect to their external momenta $\vec p_i$s. Somehow, it is rather difficult to get its solution, and so they usually take a specific limit to solve the equation. 

The reasons why we look at the conformal correlation functions in momentum space are maybe the followings. One of them is that when one develop the correlation functions in momentum space, it is very clear how the momentum flows in the interaction vertices. One may construct or say analyze the correlation functions by a Feynman-like diagrammatic language. One may also develop a bootstrap or conformal block techniques in momentum space. The second reason could be holography. To understand conformal correlation functions, rather than directly solving the conformal Wanrd identites, maybe it would better emplying holography. It is hard to specify which conformal field theory corresponds to which gravity model in general though.

In this respect, there has been several attempts to study conformally coupled scalar theory in AdS space and its field theory dual\cite{Oh:2012bx,Jatkar:2013uga,Oh:2014nfa,Papadimitriou:2007sj,deHaro:2006ymc}
in the holographic framework. In \cite{Oh:2012bx,Jatkar:2013uga,Oh:2014nfa}, the authors obtain
the dual field theory correlation functions of certain composite operators in it by employing conformally coupled scalar theories in AdS$_{d+1}$. In \cite{Oh:2012bx,Jatkar:2013uga}, they concentrate on the 2-point correlation functions of a composite-single trace operator, $O_{\Delta}$ with its conformal dimension $\Delta=\frac{d+1}{2}$ in $d$-dimensional conformal field theory. In these researches, they do not consider any interactions higher than 2-point. The conformally coupled scalar theory contains its self-interaction term in it, which has a form of $\mathcal L_{int}=\frac{\lambda}{4}\phi^\frac{2(d+1)}{d-1}$, where the $\phi$ is the scalar field and the $\lambda$ is the self-interaction coupling. When $d=3$ it becomes $\phi^4$ interaction and if $d=5$, it is $\phi^3$ interaction. In the other cases, the exponent of the self-interaction term becomes fractional. 

Holography of conformally coupled scalar model taking into account the interaction is considered and
it turns out that the results are matched with some of the results in conformal field theory computations \cite{Bzowski:2013sza,Bzowski:2019kwd,Maglio:2019grh,Oh:2020izq}. 
In the field theory computations, the conformal correlation functions are obtained by directly solving conformal Ward identities\cite{Bzowski:2013sza,Bzowski:2019kwd,Maglio:2019grh,Oh:2020izq}.
In \cite{Bzowski:2013sza}, the authors get 2- and 3- point functions of operators with arbitrary scaling dimension sitting in arbitrary (Euclidean)spactial dimension. They compute correlation functions of mixtures of scalar, vector and (rank-2) tensoral operators. The form of the 2-point correlation functions are manifest. The 3-point function is a form of integration, which is so called triple-K integral meaning that its integrand contains a product of three Bessel-K functions. In \cite{Bzowski:2019kwd}, the authors suggest $n$-point conformal correlation function and it is a complex integral form. They mostly concentrate on 4-point function and introduce cross-ratios in momentum space. If the cross ratio is a specific form, then one may interpret the integration as 3-loop integral with certain propagators. 

In \cite{Oh:2020izq}, the authors propose a fractional form of conformal $n$-point of scalar operators sharing the same scaling dimension $\Delta=\frac{d+1}{2}$ in $d$-dimensional Euclidean space. The exact form of the $n$-point function is $\langle O_1O_2...O_{n-1}O_n \rangle \sim \left(\sum_{i=1}^n|p_i|\right)^{-\alpha}$, where the $|p_i|=\sqrt{\vec p_i \cdot \vec p_i}$ is the absolute value of the $i$-th momentum coming into the vertex and $p_n=-\sum_{i=1}^{n-1} p_i$ due to momentum conservation. The $\alpha$ is a certain real number to be fixed by conformal Ward identities.

The conformally coupled scalar theory defined in $AdS_{d+1}$ space provides non-normalizable excitations which correspond to source terms of dual operators whose scaling dimension $\Delta=\frac{d+1}{2}$. The holographic 2-point, 3-point and 4-point functions computed by employing this model in AdS$_{d+1}$, AdS$_6$ and AdS$_4$ respectively present exactly the same forms of the conformal correlation functions suggested in \cite{Oh:2020izq}. In \cite{Oh:2014nfa}, the authors consider the self-interaction terms in $d=3$ case, and they compute 2-, 4- and 6- point functions of an operator $O_{\Delta=2}$ deforming the dual field theory. It turns out that the 2- and 4- point functions of the operator satisfy conformal Ward identities\cite{Oh:2020izq}

Now the frontier in this research field is getting multi-point functions beyond these. 
In this note, we explore the conformally coupled scalar theory in AdS$_{6}$ and compute holographic multi-point functions. In Sect. \ref{Solutions of a conformal scalar field action}, we explore conformally coupled scalar theory in AdS$_{6}$ and the solution of its equation of motion. 
We study this model in a new field frame where we use a field redefinition $\phi(x)=r^{\frac{d-1}{2}}f(x)$, where $\phi(x)$ is the original conformally coupled scalar field and the $f(x)$ is the field in the new field frame. The new field is effectively defined in flat Euclidean space whereas the $\phi(x)$ is defined in AdS space.

In sect.\ref{Solutions of a conformal scalar field action}, we solve the equation of motion by power expansion order by order in its coupling $\lambda$ and we get the explicit solutions upto order of $\lambda^3$. We also propose a diagrammatic way to get solution of order $n$ in $\lambda$ for the generic $n$. 
It shows a certain pattern, by which one can construct $n$-th order solution in $\lambda$ graphically.


In sect.\ref{The boundary on-shell action}, we list the holographic 2-,3- and 4-point functions explicitly from the given solutions in sect.\ref{Solutions of a conformal scalar field action}.  In this section, we discuss the details for computing the boundary on-shell action as regularity and boundary conditions at the poincare horizon and AdS boundary respectively and the counter terms for the holographic calcualtions. 

In sect.\ref{yes}, we develop the holographic Wilsonian renormalization group equations and find thier fixed points. The fixed points address the holographic correlation functions when their boundary condition is Dirichlet.

In sect.\ref{Holographic 4-point function and conformal Ward identities}, we examine the holographic 4-point function if it is really conformal. The form of the holographic $n$-point correlation functions are a function of absolute values of the external momenta $p_i$s or their linear combinations. Therefore, they are manifestly $SO(5)$ invariant, which is the rotation group in 5-dimensional Euclidean space. The rests to test are if they satisfy dilatation and special conformal Ward identities. The holographic 4-point function satisfies the dilatation Ward identity but it does not the special conformal Ward identity. 

The special conformal Ward identity has a form of ${D^\kappa{\left(p_i,\frac{\partial}{\partial p_i}\right)}}{ \Psi(p_j)}=0$, where the $D^\kappa$ is a second order differential operator with respect to the momenta $p_i$s, $\kappa$ is a spatial index and $\Psi$ is conformal $n$-point correlation function. For the holographic 4-point function that we get from the conformally coupled scalar theory in AdS$_6$, an application of $D^\kappa$ on it gives ${D^\kappa\left({p_i,\frac{\partial}{\partial p_i}}\right)}{ \Phi(p_j)}=f^\kappa(p_j)$. The $f^\kappa(p_j)$ is a function of momenta $p_i$ and it vanishes when all of the external momenta $p_i$ are co-linear, which means that $\sum_j|p_j|=|\sum_jp_j|$. This co-linear limit is interesting if we compare the holographic correlation functions of the conformally coupled scalar theory in AdS space(especially AdS$_{6}$ and AdS$_4$) with the conformal correlation functions given in \cite{Oh:2020izq}. We remark that the co-linear limit of these two correlation functions coincide.
We finally examine $n$-point holographic correlation functions, and it turns out that in the colinear limit, the conformal correlation functions given in \cite{Oh:2020izq} coincide them. Therefore, we conclude that the conformally coupled scalar theory reproduces conformal $n$-point functions in the co-linear limit but in general they are not conformal.

%



\section{Solutions of a conformal scalar field action}
\label{Solutions of a conformal scalar field action}

We start with the conformally scalar field theory defined in (Euclidean)AdS spacetime, given by
\begin{equation}
\label{con-scalar-action}
S=\int_{r>\epsilon} dr d^dx \sqrt{g} \mathcal{L}(\phi ,\partial\phi)+S_B,
\end{equation}
where the spacetime is descrived by $d+1$-dimensional Euclidean AdS metric as follows:
\begin{equation}
\label{Euc-metric}
ds^2=g_{MN}dx^M dx^N=\frac{1}{r^2} \left( dr^2+\sum_{i=1}^{d}dx^i dx^i \right)
\end{equation}
The $S_B$ is a collection of boundary terms which is designed for a well defined variational problem of the theory as we will see.
The conformally coupled scalar field Lagrangian denstiy is given by
\begin{equation}
\label{Scalar-Lagrangian}
\mathcal L(\phi,\partial \phi)=\frac{1}{2}g^{MN}\partial_M\phi \partial_N\phi+\frac{1}{2}m^2 \phi^2 +\frac{\lambda}{4}\phi^\frac{2(d+1)}{d-1},
\end{equation}
where the Latin(capital) indices $M$, $N$... run from 1 to $d+1$, where the coordinate $x^{d+1}$ denotes the AdS radial variable $r$. The mass of the conformally coupled scalar is not arbitrary, which should be
\begin{equation}
\label{mass-condition}
m^2=-\frac{d^2-1}{4},
\end{equation}
where the mass term is originated from the background curvature scalar of the AdS space. Such a value of the scalar field mass implies two different properties of the theory. One is the alternative quantization scheme in holography. In AdS/CFT context, a non-normalizable excitations in dual gravity corresponds to a source term couples to a composite operator, $\mathcal O$ in the dual field theory defined on the AdS boundary and the $\mathcal O$ is proportional to the coefficient of a normalizable mode of excitations. One of the conditions that the operator $\mathcal O$ satisfies is that its correlation functions are unitary. It turns out that it does in its mass range of
\begin{equation}
\label{alt-condition}
-\frac{d^2}{4}\le m^2 \le -\frac{d^2}{4}+1,
\end{equation}
so  does the conformally coupled scalar theory.
Moreover, if the scalar field mass is in that range, the role of normalizable and non-normalizable excitations can be switched and the boundary field theory is a still unitary theory: alternative quantization.

Another property is that by employing a scale transformation of the scalar field as
\begin{equation}
\label{con-transform}
\phi(x^M)\equiv\Omega(r)f(x^M) {,\rm \ where\ } \Omega(r)=r^{\frac{d-1}{2}},
\end{equation}
the conformally coupled scalar field action is effectively defined in a flat background as follows:
\begin{eqnarray}
\label{trans-action}
S&=&\int_{r>\epsilon}drd^dx \left( \frac{1}{2}\delta^{MN}\partial_M f(x) \partial_N f(x)+\frac{\lambda}{4}f^\frac{2(d+1)}{d-1}(x) \right) \\ \nonumber
&+&\frac{d-1}{2}\int d^dx\left. \frac{f^2(x)}{2r} \right\vert_{\epsilon}^\infty+S_B
\end{eqnarray}
The action with a new field $f$ is a massless scalar field theory with a peculiar self interaction $\sim f^{\frac{2(d+1)}{d-1}}$. The power of the self interaction becomes an integral number when $d=3$ or $d=5$, corresponding to $f^4$ and $f^3$ self interaction respectively. We may study $f^3$ case extensively.

\subsection{Small coupling expansion}
In this subsection, we will obtain the equation of motion of the conformally coupled scalar. The equation of motion is given by
\begin{equation}
\label{bulk-eom}
\delta^{MN}\partial_M\partial_Nf(x)-\frac{\lambda(d+1)}{2(d-1)}f^{\frac{d+3}{d-1}}(x)=0,
\end{equation}
where we get the equation of motion in the new field frame.
Since we are interested in the case of the self-interaction with an integral power, we choose $d=5$, namely $f^3$-interaction case in this note
\footnote{The $f^4(x)$ case is also extensively studied in \cite{Oh:2014nfa}.}
. In this case, the scalar field $\phi$ is defined in AdS$_6$ space time. The equation of motion becomes
\begin{equation}
\label{d5-case}
\left(\partial_r^2+\sum_{i=1}^{5}\partial_{i}^2\right)f(x)-\frac{3\lambda}{4}f^2(x)=0.
\end{equation}
Since the boundary directions described by the coordinates $\{x^i\}$ are flat and non-compact, we expect that the solution is a superpositon of plane waves along the boundary directions. Then, we try a solution as
\begin{equation}
\label{moment-field}
f(r,x^i)=\int\frac{d^5k}{{(2\pi)}^\frac{5}{2}} e^{-ik \cdot x}f_k(r),
\end{equation}
which is an effectively Fourier transform to the momentum space along the boundary directions.
The equation of motion in the momentum space is given by
\begin{equation}
\label{mfield-eom}
(\partial_r^2-{|q|}^2)f_q(r)-\frac{3\lambda}{4}\int\frac{d^5k}{{(2\pi)}^\frac{5}{2}}f_k(r)f_{q-k}(r)=0,
\end{equation}
and we solve this equation by using power expansion order by order in the small coupling $\lambda$ as
\begin{equation}
f_q(r)=\sum_{i=0}^\infty f^{(i)}(r),
\end{equation}
where $f^{(i)}(r)$ is $i$-th order solution in the small coupling, $\lambda$ expansion. In each order in $\lambda$, the equation of motion is given by
\begin{equation}
\label{eqobyo}
(\partial_r^2-{|q|}^2)f^{(n)}_q(r)-\frac{3\lambda}{4}\int\frac{d^5k}{{(2\pi)}^\frac{5}{2}}\sum_{i=0}^{n-1}f^{(i)}_k(r)f^{(n-i)}_{q-k}(r)=0,
\end{equation}
where we call the last terms in the equation source terms of the equation. The inhomogeneous part of solution comes from the source terms 

The zeroth order equation of motion and its solution in the expansion are given by
\begin{equation}
\label{zero-odr}
(\partial_r^2-q^2)f_q^{(0)}(r)=0 \qquad   \longrightarrow \qquad f_q^{(0)}(r)=( F_qe^{-|q|r}+ g_qe^{|q|r}),
\end{equation}
where $F_q$ and $g_q$ are arbitrary boundary momentum dependent functions.
Since the solution being proportional to $e^{|q|r}$ is divergent as $r\rightarrow\infty$, which may cause infinite energy and momentum density and so it cause huge backreaction to the background space time. To avoid this(regularity condition), we set $g_q$ to zero.
To get the higher order solution, we take the following form of the trial solution:
\begin{equation}
\label{first-odr-form}
f_q(r)=f_q^{(0)}(r)+f_q^{(1)}(r)+O(\lambda^2)
\end{equation}
The equation of motion up to the first order in $\lambda$ is given by
\begin{eqnarray}
\label{first-eom}
(\partial_r^2-q^2)f_q^{(1)}(r)&=&\frac{3\lambda}{4}\int\frac{d^5k}{{(2\pi)}^{\frac{5}{2}}}f_k^{(0)}(r)f_{q-k}^{(0)}(r) \\ \nonumber
&=&\frac{3\lambda}{4}\int \frac{d^5k}{{(2\pi)}^{\frac{5}{2}}}F_k F_{q-k} e^{-(|k|+|q-k|)r},
\end{eqnarray}
and its solution is
\begin{equation}
\label{first-odr}
f_q^{(1)}(r)=\frac{3\lambda}{4{(2\pi)}^{\frac{5}{2}}}\int d^5k d^5p {\ } \delta^{(5)}(k+p-q)\Delta(k,p) F_k F_{p} e^{-(|k|+|p|)r}+F_1(q)e^{-|q|r}, 
\end{equation}
where
\begin{equation}
\label{zero-boundary-pro}
\Delta(p_1,p_2,...,p_n)\equiv\frac{1}{(\sum_{i=1}^n|p_i|)^2-(\sum_{i=1}^np_i)^2}.
\end{equation}
The $F_1(q)e^{-|q|r}$ is homogeneous solution of the first order equation of motion, namely it satisfies $(\partial_r^2-q^2)f_q^{(1)}(r)=0$. The coefficient $F_1(q)$ is an arbitrary function of the boundary momentum $q$, but there are several ways to determine it. 
\begin{itemize}
\item One can simply choose it to vanish. The generic form of the solution in $n$-th order in $\lambda$ is given by
\begin{equation}
f^{(n)}_q(r)=f^{(n)}_{q, \rm inhomogeneous}(r)+F_n{(q)}e^{-|q|r},
\end{equation}
it means that all the $F_n(q)=0$ for $n>0$. We name such solutions $\bf Prime\  solutions(diagrams)$ for later use.
\label{vanish}
\item Another choice is that we choose it in such a way that it removes co-linear divergences in the solution. 
\end{itemize}

\begin{figure}[b!]
\centering
\includegraphics[width=200.4mm]{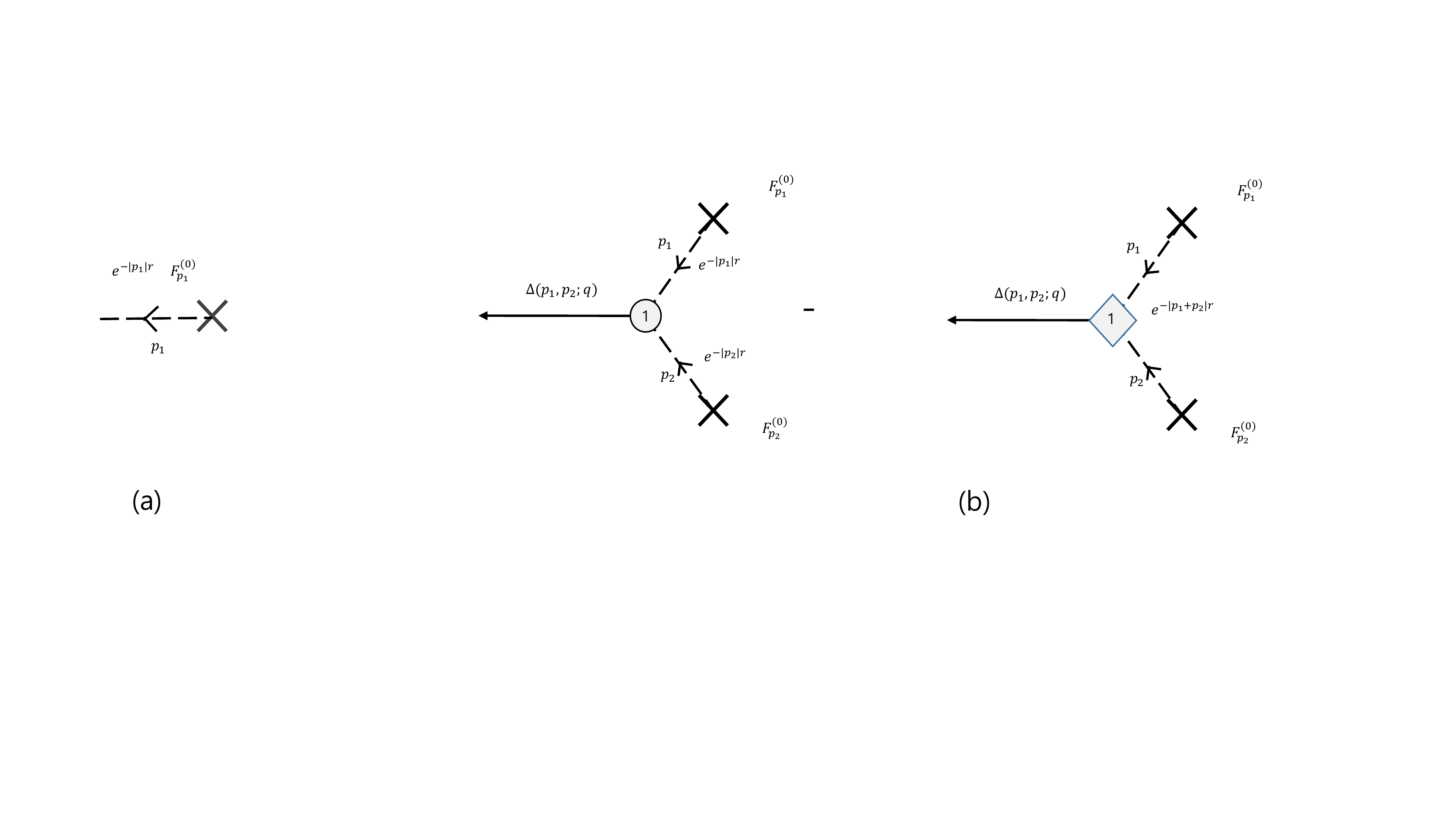}
\caption{The zeroth and the first order solutions in $\lambda$}
\label{fig2}
\end{figure}

\begin{figure}[b!]
\centering
\includegraphics[width=120.2mm]{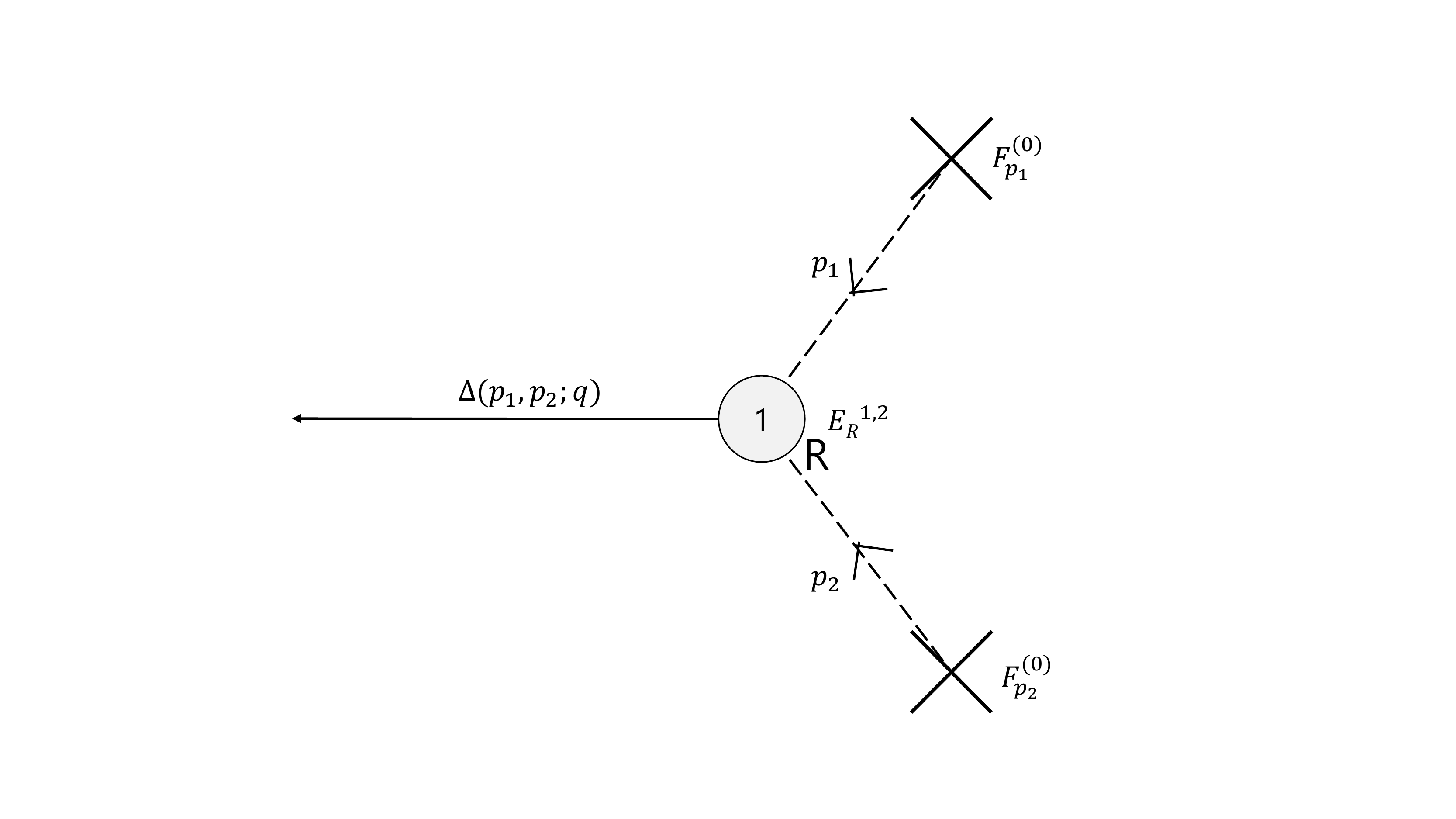}
\caption{The regularized diagram for the first order solution in $\lambda$}
\label{fig22}
\end{figure}

The factor $\Delta(p_1,p_2,...,p_n)$ diverges when
\begin{equation}
\left|\sum_{i=1}^np_i\right|=\sum_{i=1}^n|p_i|,
\end{equation}
namely, it diverges when the directions of the boundary momenta are all the same. One can remove this divergence by a smart choice of the coefficient $F_1(q)$ as
\begin{equation}
F_1(q)=-\frac{3\lambda}{4{(2\pi)}^{\frac{5}{2}}}\int d^5k d^5p {\ } \delta^{(5)}(k+p-q)\Delta(k,p) F_k F_{p}
\end{equation}
Now, the first order solution becomes
\begin{equation}
\label{FIRST-ORDER-SOLLL}
=\lambda_5 \left(\prod_{i=1}^2\int d^5p_i F_{p_i}\right)\Delta_2(p_1,p_2;q)\left\{e^{-\sum_{i=1}^2|p_i|r}-e^{-\left|\sum_{i=1}^2p_i\right|r)}\right\}
\end{equation}
where
\begin{equation}
\label{zero-boundary-pro-general}
\Delta_n(p_1,p_2,...,p_n;q)\equiv\frac{\delta^{(5)}(\sum_{i=1}^n p_i-q)}
{(\sum_{i=1}^n|p_i|)^2-q^2}.
\end{equation}
This first order solution is expressed in a diagram in Fig\ref{fig2}-(b). One may define a new symbolic variable $E_R^{1,2}$ as
\begin{equation}
E_R^{1,2}=e^{1,2}-e^{12}\equiv \left\{e^{-\sum_{i=1}^2|p_i|r}-e^{-\left|\sum_{i=1}^2p_i\right|r)}\right\},
\end{equation}
where the subscript `$_R$' denotes ``regularized''. 
For the future usage, we define 
\begin{eqnarray}
E_R^{12...m,m+1m+2...n,...,q+1...s}&=&e^{12...m,m+1m+2...n,...,q+1...s}-e^{123...s} \\ \nonumber
&\equiv& \exp\left\{ -\left(  \left|\sum_{i=1}^m p_i\right|  + 
\left|\sum_{j=m+1}^n p_j\right|+...+\left|\sum_{k=q+1}^s p_k\right| 
\right)r \right\}- \exp\left\{ -  \left|\sum_{i=1}^s p_i\right|r \right\}
\end{eqnarray}
We express the regularized diagram in Fig\ref{fig22}.

We keep performing such a process and get higher order solutions in the small coupling $\lambda$. The second order equation and solution in $\lambda$ are more complex. By using the following ansatz:
\begin{equation}
\label{second-odr-form}
f_q(r)=f_q^{(0)}(r)+f_q^{(1)}(r)+f_q^{(2)}(r)+O(\lambda^3),
\end{equation}
we get the second order equation as
\begin{eqnarray}
\label{second-eom}
\nonumber
(\partial_r^2-q^2)f_q^{(2)}(r)&=&\frac{3\lambda}{4}\int\frac{d^5k}{{(2\pi)}^{\frac{5}{2}}}\left[f_k^{(1)}(r)f_{q-k}^{(0)}(r)+f_k^{(0)}f_{q-k}^{(1)}(r)\right] =
2\cdot\frac{3\lambda}{4}\int\frac{d^5k}{{(2\pi)}^{\frac{5}{2}}}\left[f_k^{(1)}(r)f_{q-k}^{(0)}(r)\right]
\\ \nonumber
&=& 2\cdot{\left(\frac{3\lambda}{4}\right)}^2 \int\frac{d^5kd^5p}{{(2\pi)}^5}\frac{F_p^{}F_{k-p}^{}F_{q-k}^{}}{{(|p|+|k-p|)}^2-{|k|}^2}\{e^{-(|p|+|k-p|)\cdot r} - e^{-|k|\cdot r}\}e^{-|q-k|r}\\ \nonumber
\end{eqnarray}
One may realize that the first and the second terms are the same once one redefines the momenta in the second term as 
$k\equiv q-k'$ and $p\equiv p'$ in the first line in (\ref{second-eom}).
The solution of the second order equation is given by
\begin{eqnarray}
\label{second-odr}
\nonumber
f_q^{(2)}(r)&=&2\cdot{\left(\frac{3\lambda}{4}\right)}^2 \int \frac{d^5k}{{(2\pi)}^5}\left(  \prod_{i=1}^2\int d^5 p_iF_{p_i}\right)\Delta_2(p_1,p_2;k)F_{q-k}
\left\{
 \frac{e^{-(|p_1|+|p_2|+|q-k|) r}-e^{-|q|r}}{{[(|p_1|+|p_2|+|q-k|)}^2-{|q|}^2]} \right.\\ \nonumber
&-& \left.\frac{e^{-(|k|+|q-k|)r}-e^{-|q|r}}{{[(|k|+|q-k|)}^2-{|q|}^2]} \right\}, \\ \nonumber
&=&2\cdot  \left(\prod_{i=1}^3\int d^5p_i F_{p_i}
\right) (\lambda_5)^2\cdot\delta^{(5)}\left(\sum_{i=1}^3 p_i-q\right)\Delta(p_1,p_2)
\{\Delta(p_1,p_2,p_3)E_R^{1,2,3}  \\ \nonumber
&-&\Delta(p_1+p_2,p_3)E_R^{12,3}\}, 
\\ \nonumber
&=&2\cdot  \left(\prod_{i=1}^3\int d^5p_i F_{p_i}
\right) (\lambda_5)^2\cdot\int d^5k_1\Delta_2(p_1,p_2;k_1)\{\Delta_3(p_1,p_2,p_3;q)E_R^{1,2,3}\\ \nonumber
&-&\Delta_2(p_1+p_2,p_3;q)E_R^{12,3}\}
\end{eqnarray}


As the last explicit example, we discuss the the third order equation of motion and the solution of it.  To get this, we let
\begin{equation}
\label{third-odr-form}
f_q(r)=f_q^{(0)}(r)+f_q^{(1)}(r)+f_q^{(2)}(r)+f_q^{(3)}(r)+O(\lambda^4).
\end{equation}
The third order equation of motion is
\begin{equation}
\label{third-eom}
(\partial_r^2-q^2)f_q^{(3)}(r)=\frac{3\lambda}{4}\int\frac{d^5k}{{(2\pi)}^{\frac{5}{2}}}\left[2\cdot f_k^{(2)}(r)f_{q-k}^{(0)}(r)+f_k^{(1)}(r)f_{q-k}^{(1)}(r)\right],
\end{equation}
%
%
%
and the third order solution is
\begin{eqnarray}
\label{third-odr}
\nonumber
f_q^{(3)}(r)
&=& (\lambda_5)^3\left(\prod_{i=1}^4\int d^5p_i F_{p_i}\right) \cdot\left(\prod_{j=1}^2\int d^5k_j\right)[ \Delta_2(p_1,p_2;k_1)\Delta_2(p_3,p_4;k_2)\{\Delta_4(p_1,p_2,p_3,p_4;q)E_R^{1,2,3,4} \\ \nonumber
&-&2\Delta_3(p_1,p_2,p_3+p_4)E_R^{1,2,34}+\Delta_2(p_1+p_2,p_3+p_4)E^{12,34}_R\} \\ \nonumber
&+&2^2\Delta_2(p_1,p_2;k_1)\Delta_3(p_1,p_2,p_3;k_2)\{\Delta_4(p_1,p_2,p_3,p_4;q)E^{1,2,3,4}_R-\Delta_2(p_1+p_2+p_3,p_4)E^{123,4}_R\} \\ \nonumber
&-&2^2\Delta_2(p_1,p_2;k_1)\Delta_2(p_1+p_2,p_3;k_2)\{\Delta_3(p_1+p_2,p_3,p_4;q)E^{12,3,4}_R-\Delta_2(p_1+p_2+p_3,p_4)E^{123,4}_R\}] 
\end{eqnarray}



\begin{figure}[b!]
\centering
\includegraphics[width=128.4mm]{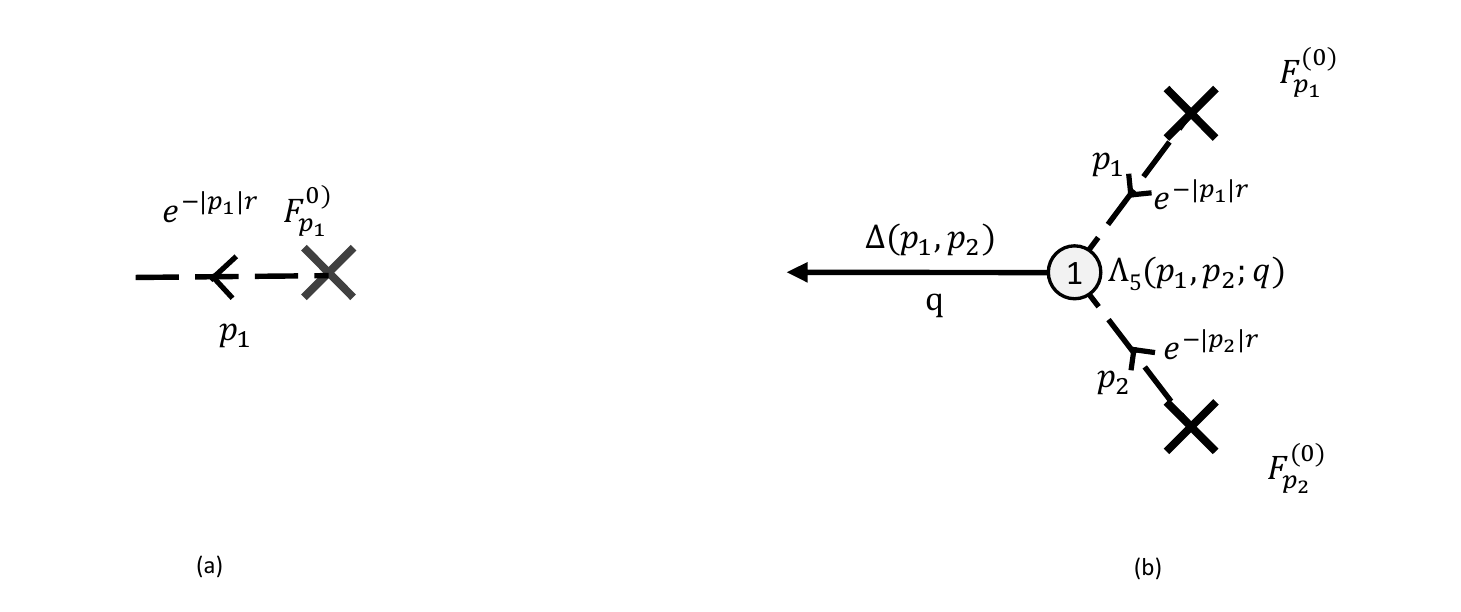}
\caption{The prime diagrams for the zeroth(a) and the first9b) order solutions in $\lambda$}
\label{rrE1}
\end{figure}

\begin{figure}[b!]
\centering
\includegraphics[width=60.8mm]{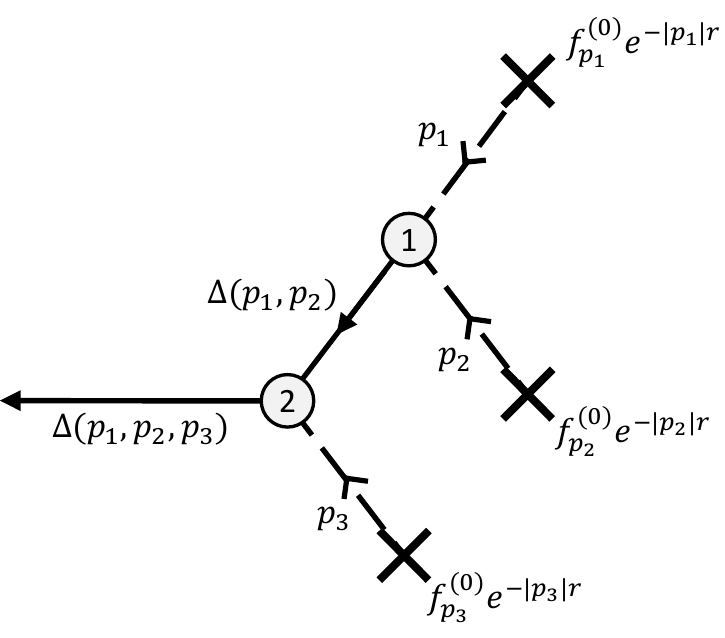}
\caption{The prime diagram for the second order solution in $\lambda$ }
\label{fig32}
\end{figure}

\begin{figure}[b!]
\centering
\includegraphics[width=130.07mm]{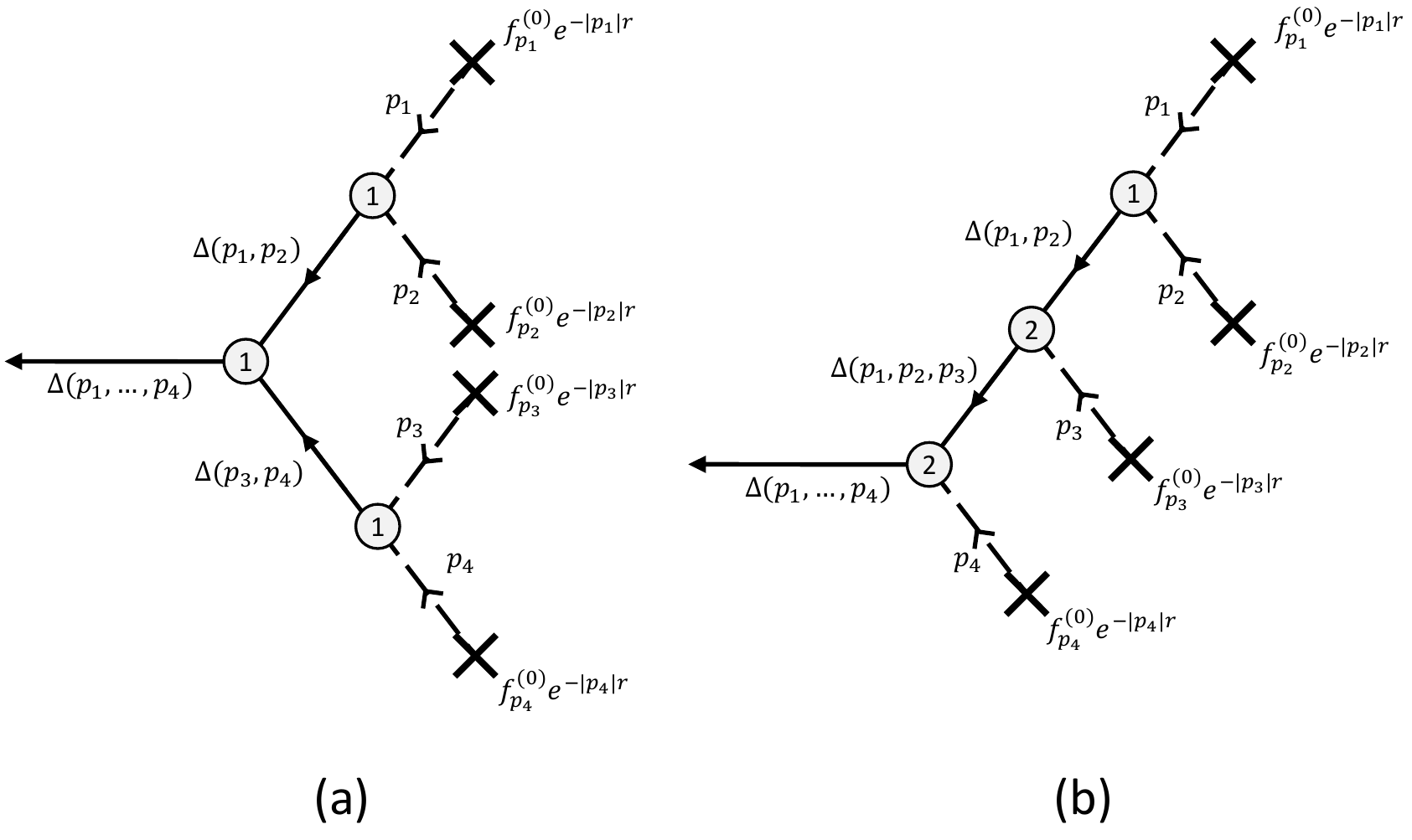}
\caption{The prime diagram for the third order solutions in $\lambda$}
\label{fig3}
\end{figure}

\begin{figure}[b!]
\centering
\includegraphics[width=170mm]{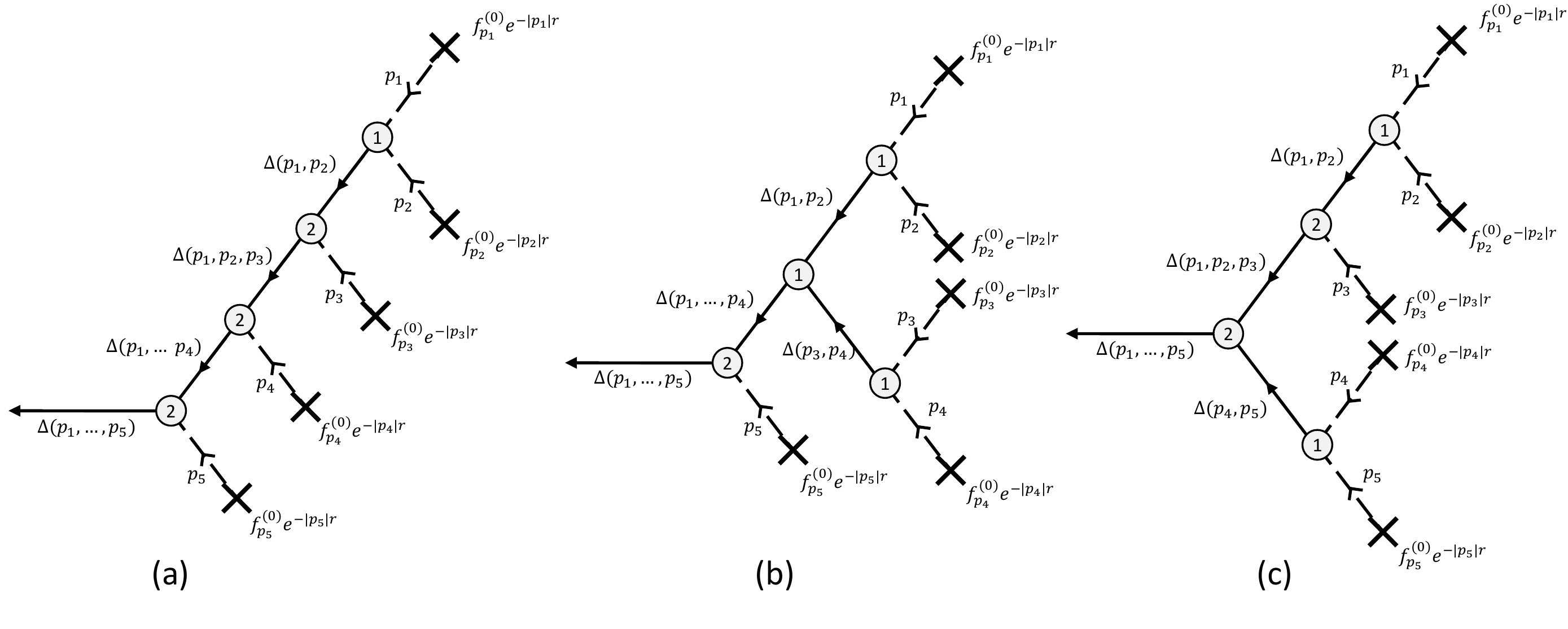}
\caption{The prime diagram for the fourth order solutions in $\lambda$ }
\label{fig4}
\end{figure}

\begin{figure}[b!]
\centering
\includegraphics[width=122.07mm]{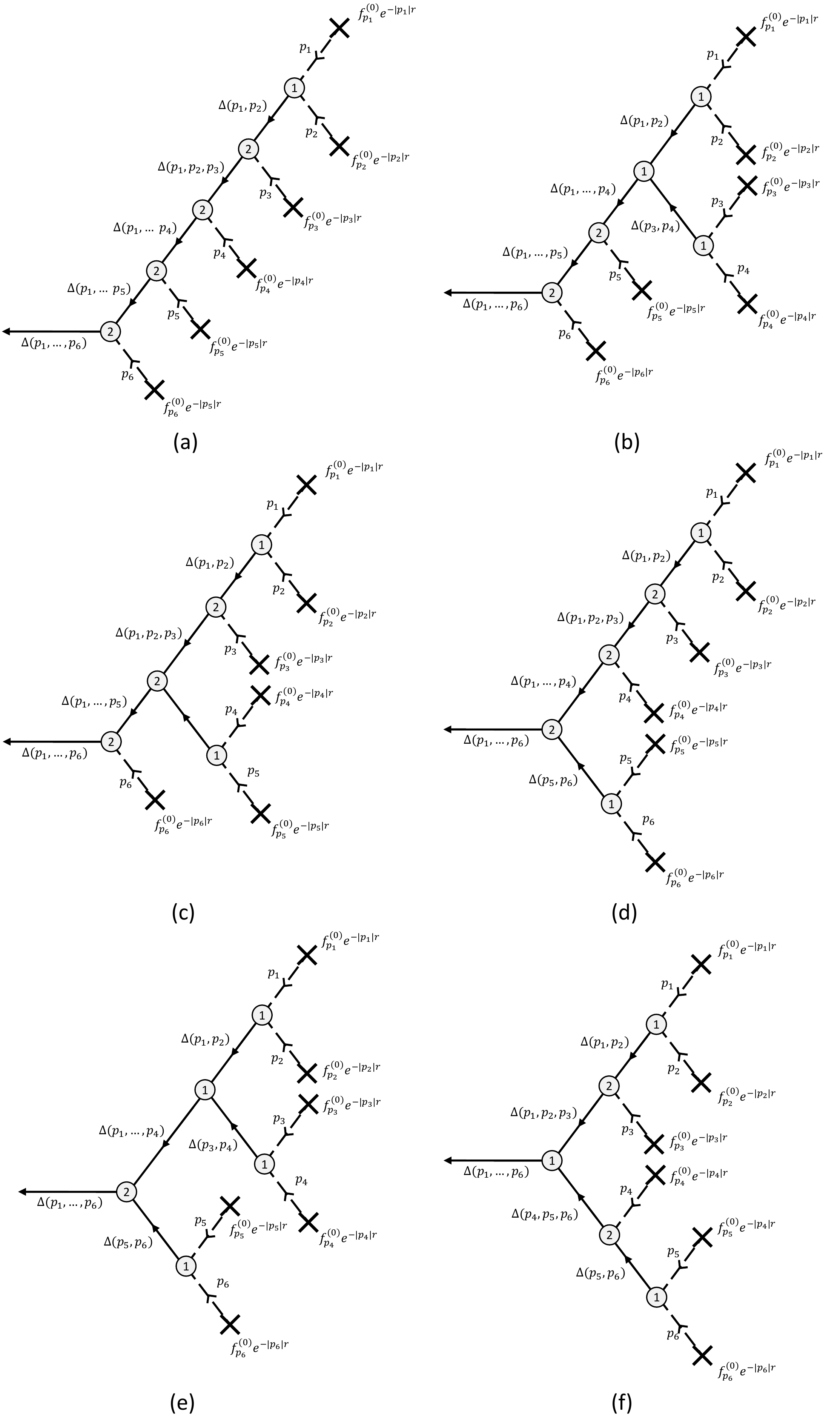}
\caption{The prime diagram for the fifth order solutions in $\lambda$ }
\label{fig5}
\end{figure}

\paragraph{A pictorial way to express the solution in the small coupling expansion}
Even if perturbative solutions become more complex as they become higher order, the solutions that we obtain 
show a certain pattern. Once we observe the solutions in detail, 
we realize that we can express the solutions with a collection of certain diagrammatic elements. 


To formulate each pictorial element, let us concentrate on the $\bf prime\ diagrams$ the we defined previously. Again, the prime diagrams are the solutions of the equation of motion(\ref{eqobyo}) with all the $F^{(n)}_i(p)=0$ except $n=0$ case, which are the coefficients of the homogeneous solutions in each order in power expansion in $\lambda$.

Let us look at the zeroth order solution(\ref{zero-odr}). The solution is $f_q^{(0)}(r)= F_q^{(0)}e^{-|q|r}$, and its near AdS boundary expansion is
\begin{equation}
f_q^{(0)}(r)= F_q^{(0)}-r|q|F_q^{(0)}+O(r^2){\rm \ \ \ as\ \ }r\rightarrow0,
\end{equation}
where the $F_q^{(0)}$ is the boundary value of the solution and it corresponds to a coefficient of the non-normalizable mode of the excitation(solutions). In  the context of AdS/CFT correspondence, this coefficient corresponds to a boundary source coupled to a composite operator in the dual conformal field theory. 

The factor $e^{-|q|r}$ is an evolution of the boundary value $F_q^{(0)}$ to the interior of AdS space along the its radial direction. Namely, it is the radial propagation of the solution. We express the boundary source $F_q^{(0)}$ by small cross and the radial propagator by dotted lines. The zeroth order solution is expressed at (a) in Figure.\ref{fig2}.

The first order solution in $\lambda$, $f^{(1)}_q$ given in (\ref{first-odr}),  is a little more complex than the zeroth order one. From the two different boundary source terms, $F^{(0)}_{p_1}$ and $F^{(0)}_{p_2}$, the fields propagate into the interior of AdS space and so its amplitude contains the factor of $e^{-|p_1|r}$ and $e^{-|p_2|r}$ in it. These are expressed by crosses and dotted lines as we do for the zeroth order solution. One more thing that we need to stress is that since the momenta $p_1$ and $p_2$ can be arbitrary, $\int d^5 p_1 d^5p_2$ factor should be inserted in the amplitude.

After the radial propagation, these two fields interact each other with  a coupling, $\lambda_5\equiv\frac{3\lambda}{4(2\pi)^{\frac{5}{2}}}$ and create another (scalar) field with momentum $q$. It is a 3-point interaction vertex. 
The newly created field does not propagate in the radial direction but only along the boundary directions. The boundary propagator, $\Delta_2(p_1,p_2;q)$ is given in (\ref{zero-boundary-pro-general}) and the boundary propagator contains $\delta$-function for the momentum conservation. We express the boundary propagator by solid lines.
After all,  the first order solution can be demonstrated as given at (b) in the figure.(\ref{fig2}) in terms of the diagrammatic elements that we introduced. 

The boundary propagator is not a usual two point amplitude since it depends on $p_1$ and and $p_2$ as well as $q=p_1+p_2$. Somehow it shows a non-local structure of the amplitude and the non-locality will be manifest as it goes higher orders.

\begin{figure}[b!]
\centering
\includegraphics[width=120mm]{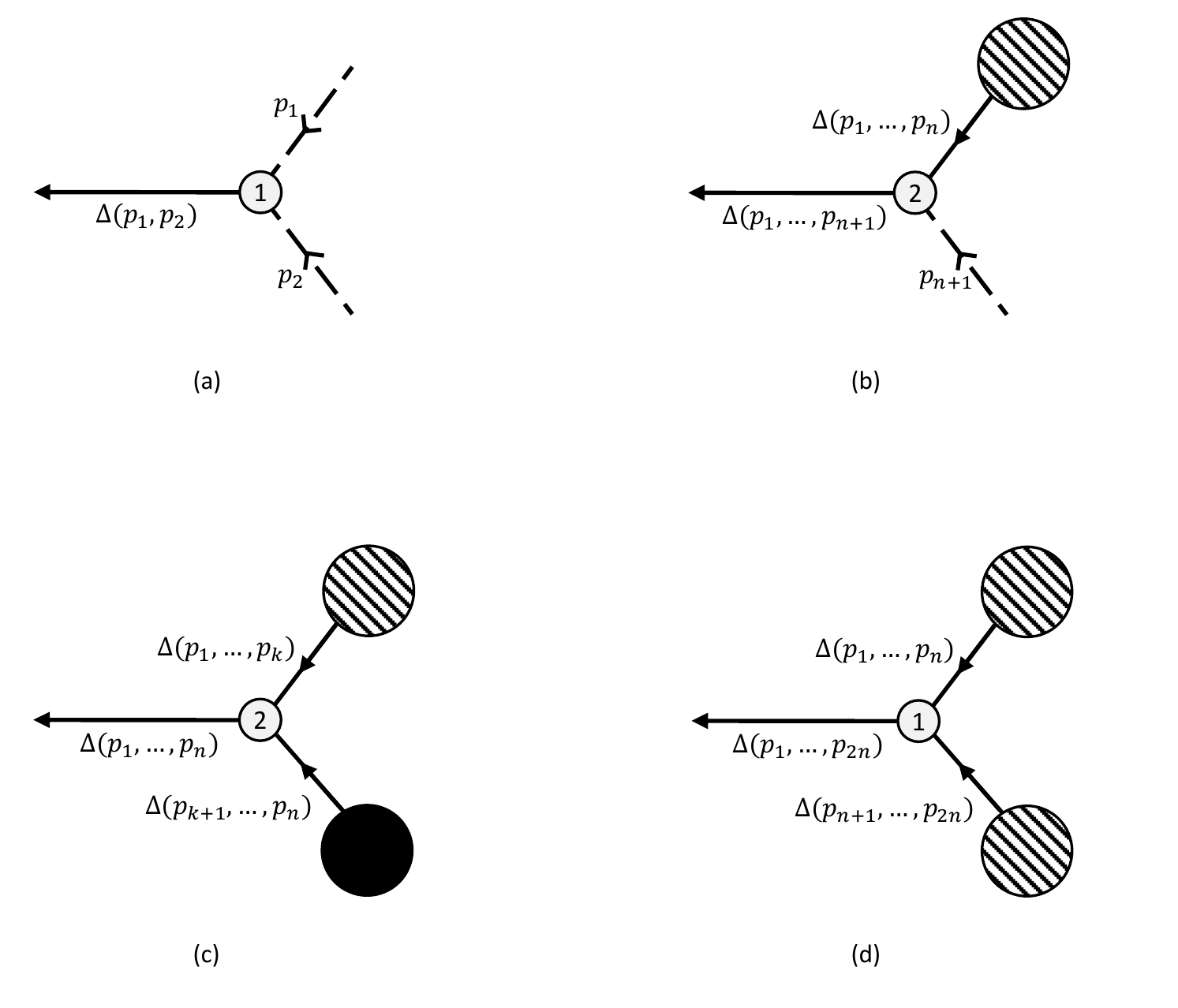}
\caption{Each diagrammatic elements}
\label{fig1}
\end{figure}

\begin{figure}[b!]
\centering
\includegraphics[width=128.4mm]{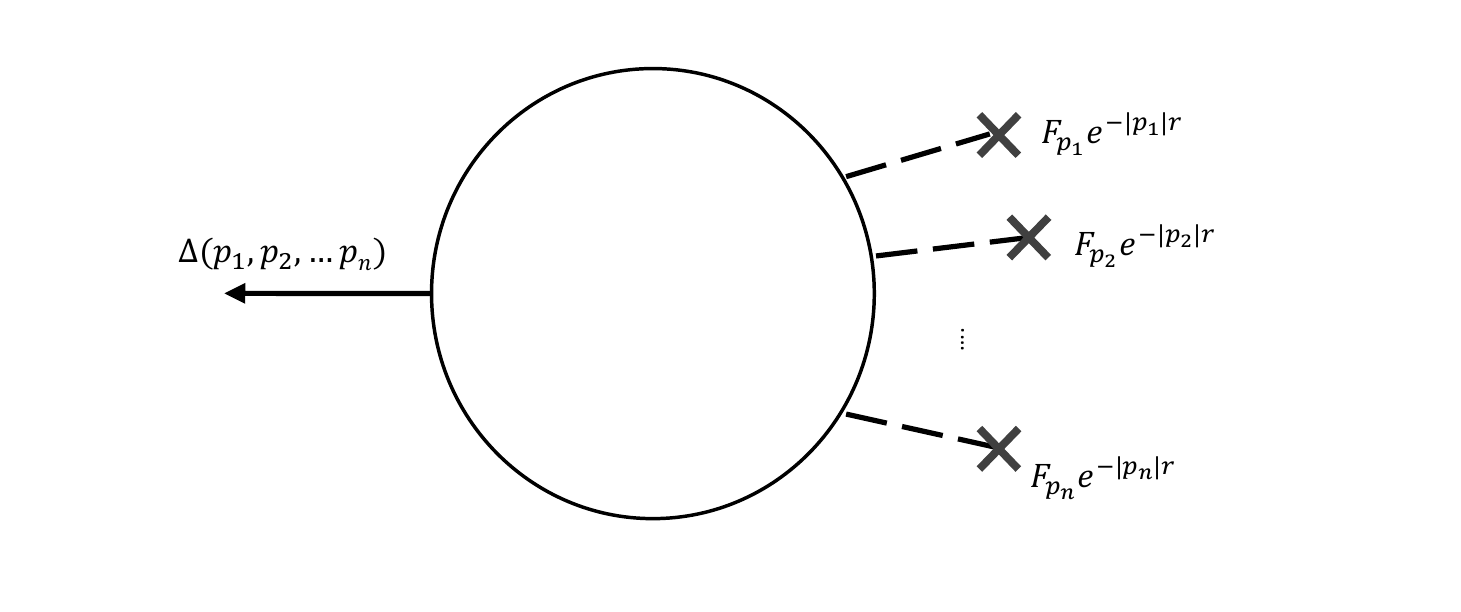}
\caption{The general construction of the classical solutions}
\label{figcollection}
\end{figure}

\paragraph{General rules}  By looking at many of the higher order solutions together with the zeroth and the first order solutions, we find the following rules. 
\begin{itemize}
\item The $n$-th order solution in the small coupling $\lambda$, $f^{(n)}(r)$ given in Fig.\ref{figcollection} contains $n+1$ boundary source terms $F_{p_i}$. All these source terms propagate in the radial direction therefore the solution contains the factor of $\prod_{i=1}^{n+1}\int d^5k_ie^{-|k_i|r}F_{k_i}$. These can be expressed in terms of collection of $n+1$ copies of the diagramatic element (but with different momentum for each dotted line) given at (a) in Fig. \ref{fig2}
\item After the radial propagation of the source terms, certain pairs of the radial propagators expressed in the dotted lines merge into a point where the propagators interact each other with a strength of the coupling constant of $\lambda_5$. This vertex produces another type of propagator, ``boundary propagator" $\Delta_2(p_1,p_2;q)$, 
which is expressed in a solid line. The solid line will interact with a radial or boundary propagetor to produce another boundary propagator $\Delta_i(p_1,p_2,...,p_i;q)$. 
\item There are 4 different types of interaction vertices given in Fig.\ref{fig1}. (a) in Fig.\ref{fig1} demonstrates that two radial propagators, $\prod_{i=1}^2\int d^5 p_iF_{p_i}e^{-|p_i|r}$ interact each other with coupling $\lambda_5$ to produce a boundary propagator  $\Delta_2(p_1,p_2;q)$. (b) in Fig.\ref{fig1} indicates that a radial propagator, $\int d^5 p_{n+1}F_{p_{n+1}}e^{-|p_{n+1}|r}$ and a boundary propagator, $\Delta_n(p_1,p_2,...,p_n;k)$ interact each other to produce another boundary propagator,  $\Delta_{n+1}(p_1,p_2,...,p_{n+1};q)$. The factor `2' inside the small circle at the interaction vertex means that we multiply a factor of ``2'' to the amplitude.
\item (c) and (d) in Fig.\ref{fig1} show that the two boundary propagators,  $\Delta_n(p_1,p_2,...,p_k;k_1)$ and  $\Delta_n(p_{k+1},p_{k+2},...,p_n;k_2)$ interact each other and produce another boundary propagator. At (c) in Fig.\ref{fig1}, the two boundary propagators coming into the vertex have different histories. In such a case, we mulpiply a factor of 2 to the amplitude. At (d) in Fig.\ref{fig1}, the two boundary propagators coming in share the same history. In this case, $n$ is an even number and $k=\frac{n}{2}$ and we muliply a factor of 1 to the amplitude. 
\item The $n$-th order solution in $\lambda$ is a collection of all possible connections of the radial propagators via the 4-different types of the interaction vertices and finally produce a single boundary propagator as seen in Fig.\ref{figcollection}. For all the internal boundary propagator, $\Delta_n(p_1,p_2,...,p_k;k_1)$,  there should an integration, $\int dk_1$ inside the amplitude.
\item {\bf Caution}: We do not multi-count any amplitudes(or diagrams) which can be obtained by any permutations of the boundary momenta ``$p_i$"s. Namely, from an amplitude, if one gets another amplitude by an operation of such a permutation on it, then they are the same diagram and so we do not double count them.
\end{itemize}

\paragraph{Recovering the full solutions}
The prime solution that we obtained from the general rules addressed is given by
\begin{equation}
f_p(r)=\sum_{l=1}^\infty \lambda_5^{l-1}\left( \prod^l_{j=1}\int d^5p_{j}F_{p_j})\right)A_l(p_1,...,p_l;p),
\end{equation}
where
\begin{eqnarray}
\label{primedtt-solutions}
A_1(p_1;p)&=&\delta^{(5)}(p_1-p)e^1,{\ \ \ }A_2(p_1,p_2;p)=\Delta_2(p_1,p_2;p)e^{1,2}, \\ \nonumber
A_3(p_1,p_2,p_3;p)&=&2\cdot\mathcal P\left\{\int d^5 k_1 \Delta_2(p_1,p_2;k_1)\Delta_3(p_1,p_2,p_3;p)e^{1,2,3}\right\} \\ \nonumber
A_4(p_1,p_2,p_3,p_4;p)&=&\mathcal P\left\{\left(\prod_{j=1}^2\int d^5k_j\right)[ \Delta_2(p_1,p_2;k_1)\Delta_2(p_3,p_4;k_2)\Delta_4(p_1,p_2,p_3,p_4;q)\right. \\ \nonumber
&+&2\left.\cdot 2{\ }\Delta_2(p_1,p_2;k_1)\Delta_3(p_1,p_2,p_3;k_2)\Delta_4(p_1,p_2,p_3,p_4;q)]e^{1,2,3,4}\right\}, \\ \nonumber
{\rm \ and\  so\  on...}
\end{eqnarray}
and the symbol $\mathcal P$ denotes all possible permutations of the momenta $p_i$ in the expressions. {\bf How can we recover the full solution? A way is to add all possible terms which cancel the colinear divergences in the prime diagrams. }
\begin{itemize}
\item $A_1$ has no divergence in it(the $\delta$-function will be integrated). Then we leave it as it is.
\item $A_2$ diverges when the $p_1$ and $p_2$ are colinear. Then, we add a term as
\begin{equation}
\Delta_2(p_1,p_2;p)e^{1,2}\rightarrow \Delta_2(p_1,p_2;p)e^{1,2}-{\bf \Delta_2(p_1,p_2;p)e^{12}}=\Delta_2(p_1,p_2;p)E_R^{1,2},
\end{equation}
where the term that we add share the factor `$\Delta_2(p_1,p_2;p)$' with a different coefficient $e^{12}$. This term simply substract the divergence when $1,2=12$, which means that $|p_1|+|p_2|=|p_1+p_2|$.
\item $A_3$ has two different types of colinear divergences. When $1,2=12$, $\Delta_2(p_1,p_2;k_1)$ diverges and when $1,2,3=123$ meaning that $|p_1|+|p_2|+|p_3|=|p_1+p_2+p_3|$, $\Delta_3(p_1,p_2,p_3;p)$ is divergent. First, we remove the colinear divergence in  
$\Delta_2(p_1,p_2;k_1)$. To do this, we perform
\begin{eqnarray}
 &{\ }&\Delta_2(p_1,p_2;k_1)\Delta_3(p_1,p_2,p_3;p)e^{1,2,3}  \\ \nonumber
&\rightarrow&
\Delta_2(p_1,p_2;k_1)\Delta_3(p_1,p_2,p_3;p)e^{1,2,3}-{\bf  \Delta_2(p_1,p_2;k_1)\Delta_2(p_1+p_2,p_3;p)e^{12,3}}
\end{eqnarray}
as we do in the $A_2$ case. 
We compose the second term as follow. We keep the factor, $\Delta_2(p_1,p_2;k_1)$ in the prime diagram and change its coefficient. The coefficient is $\Delta_3(p_1,p_2,p_3;p)e^{1,2,3}$ and then we switch the variable $|p_1|+|p_2|$ by $|p_1+p_2|$ in it. Then, the coefficient becomes $\Delta_2(p_1+p_2,p_3;p)e^{12,3}$.

However, the expression still has the divergence in $\Delta_3(p_1,p_2,p_3;p)$, when $1,2,3=123$. To regularize the divergences from each term above, we add the other terms as below.
\begin{eqnarray}
&\rightarrow&
\Delta_2(p_1,p_2;k_1)\Delta_3(p_1,p_2,p_3;p)e^{1,2,3}-{\bf \Delta_2(p_1,p_2;k_1)\Delta_3(p_1,p_2,p_3;p)e^{123}}
 \\ \nonumber
&{\ }&-\Delta_2(p_1,p_2;k_1)\Delta_2(p_1+p_2,p_3;p)e^{12,3}
+{\bf \Delta_2(p_1,p_2;k_1)\Delta_2(p_1+p_2,p_3;p)e^{123}},
\end{eqnarray}
where the second term substract the colinear divergence of the first term and the fourth one does from the third one. The second term is designed by keeping the factor $\Delta_2(p_1,p_2;k_1)\Delta_3(p_1,p_2,p_3;p)$ and change its coefficient $e^{1,2,3}$ in a way that we switch the variable $|p_1|+|p_2|+|p_3|$ by $|p_1+p_2+p_3|$.  The third term is designed by keeping the factor $\Delta_2(p_1,p_2;k_1)\Delta_2(p_1+p_2,p_3;p)$ and change its coefficient $e^{12,3}$ in a way that we switch the variable $|p_1+p_2|+|p_3|$ by $|p_1+p_2+p_3|$ i it.

Finally the full solution is given by
\begin{equation}
A_3\rightarrow\Delta_2(p_1,p_2;k_1)\Delta_3(p_1,p_2,p_3;p)E_R^{1,2,3}
-\Delta_2(p_1,p_2;k_1)\Delta_2(p_1+p_2,p_3;p)E_R^{12,3}
\end{equation}
\item $A_4$ has two different terms in it. We perform the substraction process to each term and then we get the full solutions.
\end{itemize}


I
f


\section{The boundary on-shell action}
\label{The boundary on-shell action}

\subsection{Regularity and boundary conditions}
In our setting of the AdS geometry given in (\ref{Euc-metric}), the AdS boundary is at $r=0$ and the location of $r=\infty$ is for Poincare horizon, which is the deep interior of the AdS space. 
At the zeroth order in  $\lambda$, we have the two independent solutions as
\begin{equation}
\phi_1(p_i,r)=F_pr^{\frac{d-1}{2}}e^{-|p|r} {\rm\ \ or\ \ }\phi_2(p_i,r)=g_pr^{\frac{d-1}{2}}e^{|p|r},
\end{equation}
where the coefficients $F_p$ and $g_p$ are arbitrary (boundary directional) momentum ``$p_i$'' dependent functions. The $\phi_1$ is regular everywhere but the $\phi_2$ does not. It shows divergence at $r=\infty$, the Poincare horizon unless the absolute value of the momentum $|p|$ vanishes. We stress that these solutions are entire solution for the conformally coupled scalar in zeroth order in $\lambda$, not being an approximate solutions with their near boundary expansions.

Let us argue this $r=\infty$ behavior of the solutions by computing the stress-energy tensor of conformally coupled scalar theory near the Poincare horizon. This stress energy tensor of the conformally coupled scalar theory is given by
\begin{equation}
T_{MN}=\partial_M\phi \partial_N \phi - g_{MN}\mathcal L,
\end{equation}
where the  $\mathcal L$ is the Lagrangian density of the conformally coupled scalar theory, given in (\ref{Scalar-Lagrangian}). We want to evaluate this in (boundary directional) momentum space, by Fourier transform given as (\ref{moment-field}) and which gives
\begin{eqnarray}
\nonumber
T^{MN}(p)&=&\int\frac{ d^dk}{(2\pi)^{\frac{d}{2}}}\left\{ g^{rM}g^{rN}\partial_r\phi_k \partial_r \phi_{p-k} +g^{\mu M}g^{\nu N} k_\mu(p-k)_\nu \phi_k \phi_{p-k}-\frac{g^{MN}}{2}\left( g^{\mu\nu}k_\mu(p-k)_\nu \phi_k \phi_{p-k}
\right.
\right. \\
&+&\left.\left.g^{rr}\partial_r\phi_k \partial_r \phi_{p-k} +m^2\phi_k \phi_{p-k}\right)\right\},
\end{eqnarray}
where $g^{MN}=r^2\delta^{MN}$, more precisely $g^{rr}=r^2$ and $g^{\mu M}=r^2 \delta^{\mu M}$.

Now, to eveluate the stress energy tensor, consider the general solution of $\phi$, which is given by a linear combination of the $\phi_1$ and $\phi_2$. Near the Poincare horizon, the $\phi_2$ and its derivatives are more dominant than those of $\phi_1$. Therefore, 
the solution $\phi$ becomes $\phi\rightarrow \phi_1$ as $r\rightarrow\infty$. First, let us compute $T^{rr}$.
\begin{eqnarray}
\nonumber
T^{rr}(p)&=&\int\frac{d^d k}{(2\pi)^{d/2}}\phi_2(k)\phi_2({p-k})\left\{ \frac{r^4}{2}\left(|k||p-k|-k\cdot(p-k)\right) +\frac{(d-1)}{4}r^3(|k|+|p-k|)\right. \\ 
&+&\left.\frac{d(d-1)}{4}r^2\right\}
\end{eqnarray}
We also compute
$T^{\tau\tau}$, which is given by
\begin{eqnarray}
T^{\tau\tau}(p)&=&\int\frac{d^d k}{(2\pi)^{d/2}}\phi_2(k)\phi_2({p-k})
\left\{ \frac{r^4}{2}\left( (E_p-E_k)E_k-|k||p-k|-k\cdot(p-k)\right) \right. \\ \nonumber
&+&\frac{(d-1)}{4}r^3(|k|+|p-k|)+\left.\frac{d-1}{4}r^2\right\},
\end{eqnarray}
where $E_k$ is an energy of an excitation in the boundary field theory, which can be defined in the context of radial quantization, and the $\tau$ is the radial variable.
 
In the stress energy tensor, there is a factor of $\phi_2(p)\phi_2(p-k)$ in it, which diverges near the Poincare horizon as $\phi_2(p)\phi_2(p-k)\sim \exp\left\{(|k|+|p-k|)r\right\}$ as $r\rightarrow\infty$. Therefore, this causes an infinite stress energy tensor, which implies a large back reaction to the background geometry near the Poincare horizon whereas the AdS boundary does not change. The geometry can change into a black brane(but we do not examine the system for such a case in detail). If we assume as such, we may have different kinds of states(may be thermal states) in the dual field theory corresponding to the black brane solution and the holographic correlation function must not be the same with those from pure AdS space.

In this paper, what we want to do is to compute holographic correlation functions from a model of conformally coupled scalar theory in the background of pure AdS space. Therefore, we keep the Poincare horizon as that of AdS space.

\paragraph{Divergences near AdS boundary and the counter terms}
To evaluate holographic correlation functions from gravity models, one need to regularize the boundary on-shell action near AdS boundary at $r=\epsilon$ where $\epsilon<<1$, showing divergences as the boundary approaches $\epsilon\rightarrow0$ in general.

For this, the holographic renormalization process is needed. Practically, we add counter terms in the action. In our case, it is
\begin{equation}
S_{ct}=\frac{d-1}{2}\int d^d x \sqrt{\gamma}\phi^2(x),
\end{equation}
where $\gamma$ is determinant of the $\gamma_{\mu\nu}$, which is defined as
\begin{equation}
\gamma_{\mu\nu}=\frac{\partial x^M}{\partial x^\mu}\frac{\partial x^N}{\partial x^\nu}g_{MN}
\end{equation}
at $r=\epsilon$. This counter term respects AdS boundary diffeomorphism invariance and precisely cancels the divergence pieces from the boundary on-shell action that we will compute.
One can see if this counter term works correctly in the new field frame. By employing the field transformation(\ref{con-transform}), one sees that another boundary term
appers, which is the first term in the second line in (\ref{trans-action}). The counter term precisely cancel this term. Then, in this field frame, the solutions of the equation of motion and the boundary on-shell action from them become regular manifestly.

\paragraph{Boundary conditions on the AdS boundary}
Since the mass of conformally coupled scalar is in a window as $-\frac{d^2}{4}\le m^2 \le -\frac{d^2}{4}+1$, various quantization schemes are possible. Accordingly, in the dual gravity theory, various boundary conditions are suggested for these, Dirichlet, Neumann and Mixed boundary conditions.
In our computation, we restrict ourselves imposing Dirichlet boundary condition on the AdS boundary. This is for a holographic computations of correlation functions of an operator, of which conformal dimansion $\Delta=\frac{d+1}{2}$ coupled with a source $J$.

\subsection{Computation of the boundary on-shell action}
The form of the on-shell action is given by
\begin{eqnarray}
S_{os}&\equiv&S^{(1)}_{os}+S^{(2)}_{os}=\frac{1}{2}\int \left[\prod_{i=1}^2 d^5 k_i\right]f_{k_1}(r)\partial_r f_{k_2}(r)\delta^{(5)}\left.\left(\sum_{i=1}^2 k_i\right)\right|^{r=\infty}_{r=0} \\ \nonumber
&-&\frac{\lambda_5}{6}\int^\infty_0 dr  \left[\prod_{i=1}^3 f_{k_i}(r) d^5 k_i\right]\left(\sum_{i=1}^3 k_i\right),
\end{eqnarray}
where the $f_k(r)$ is the solution that we obtaine in the previous section. We note that the solution $f_k(r)$ has no colinear divergences in it and so the on-shell action is finite too except the infra-red point where all the momenta $k_i=0$. The process computing the boundary on-shell action to get $n$-point function of the composit operator which corresponds this solution is somewhat long and tedious. We just address a few of them below. 
\begin{itemize}
\item $\langle O_\Delta(k_1) O_{\Delta}(k_2) \rangle=\frac{|k_1|}{2}\delta^{(5)}(k_1+k_2)$.
\item $\langle O_\Delta(k_1) O_{\Delta}(k_2) O_{\Delta}(k_3)\rangle=
\frac{\lambda_5}{3\cdot(\sum_{i=1}^3|k_i|)}\delta^{(5)}(k_1+k_2+k_3)$.
\item
\begin{eqnarray}
\nonumber
\langle O_\Delta(k_1) O_{\Delta}(k_2) O_{\Delta}(k_3) O_{\Delta}(k_4)\rangle=-\frac{\lambda^2_5}{6\cdot(\sum_{i=1}^4|k_i|)}\left(\frac{1}{(|k_1|+|k_2|+|k_1+k_2|)(|k_3|+|k_4|+|k_3+k_4|)}\right. \\ \nonumber
+\left.({k_1 \leftrightarrow k_3})+({k_1 \leftrightarrow k_4})\right)\delta^{(5)}(k_1+k_2+k_3+k_4)
\end{eqnarray}
\item And so on.
\end{itemize}

\section{Holographic correlation functions as fixed points of holographic Winsonain RG 
}
\label{yes}
\subsection{Holographic renormalization and Hamilton Jacobi equation}
In this section, we would like to compute holographic Wilsonian renormailization group flows of a certain 5-dimensional conformal field theory when its bulk dual is given by 6-dimensional conformally coupled scalar theory in AdS$_{6}$. When the conformally coupled scalar theory(\ref{Scalar-Lagrangian}) is defiend in the AdS$_6$, its enjoys 3-point self interaction. The action is given by
\begin{eqnarray}
 \nonumber
\label{phi3-bulk-action}
S&=&\int_{r>\epsilon} dr \left[  \frac{1}{2} \int d^5k d^5 k^\prime \delta^{(5)}(k+k^\prime)     \left( \partial_r f_k \partial_r f_{k^\prime} + k^2 f_k f_{k^\prime} \right)   +\frac{\lambda}{4(2\pi)^{5/2}}\int \prod_{i=1}^3 d^5 k_i f_{k_i}\delta^{(5)}\left(\sum_{j=1}^3 k_j\right)      \right]  \\
&+&S^\prime_B(\epsilon),
\end{eqnarray}
where $k_i$ are 5-dimensional boundary directional momenta, $S_B^\prime$ is the boundary deformation, i.e. the boundary generating functional on the $r=\epsilon$ hypersurface. We define this action in the momentum space of $k_i$ by using the fourier transform that we define in the last section.

The canonical momentum and the equation of motion 
are given by
\begin{equation}
\Pi_k\equiv\partial_r f_{-k}=\frac{\delta S^\prime_B}{\delta f_k}, {\ \ \ }\partial_r \Pi_k=k^2f_k + \frac{3\lambda}{4}\int \frac{d^5k^\prime}{(2\pi)^{5/2}}f_{k^\prime} f_{k-k^\prime},
\end{equation}
where the first equation ensures that a variation with respect to the field $f_k$ is well posed. 
The fact that the total action $S$ does not depend on the radial cut-off $\epsilon$ leads the Hamilton-Jacobi equation(HJ-equation)\cite{
Oh:2012bx,Jatkar:2013uga,Heemskerk:2010hk,Faulkner:2010jy,Oh:2013tsa,Oh:2015xva,Moon:2017btx} which is given by
\begin{eqnarray}
\partial_\epsilon S^\prime_B(\epsilon)&=&-\frac{1}{2}\int d^5 k \left( \frac{\delta S^\prime_B}{\delta f_k(\epsilon)} \right)\left( \frac{\delta S^\prime_B}{\delta f_{-k}(\epsilon)}\right) +   \frac{1}{2} \int d^5k d^5 k^\prime \delta^{(5)}(k+k^\prime)      k^2 f_k f_{k^\prime}  \\ \nonumber
 &+&\frac{\lambda}{4}\int \prod_{i=1}^3 d^5 k_i f_{k_i}\delta^{(5)}\left(\sum_{j=1}^3 k_j\right)
\end{eqnarray}
To get the solution of the equation, we suggest the following trial soltuion:
\begin{eqnarray}
S^\prime_B(\epsilon)&=&\Lambda(\epsilon)+\int J_k(\epsilon) f_{-k}(\epsilon)d^5 k +   \left[\int\prod_{i=1}^2 d^5 k_i f_{k_i}(\epsilon)\right]D^{(2)}_{k_1k_2}(\epsilon)\delta^{(5)}\left(\sum_{j=1}^2 k_j\right)\\ \nonumber
&+&\sum_{n=1}^{\infty}\lambda^n\left[\int\prod_{i=1}^{n+2} d^5 k_i f_{k_i}(\epsilon)\right]D^{(n+2)}_{k_1,...,k_{n+2}}(\epsilon)\delta^{(5)}\left(\sum_{j=1}^{n+2} k_j\right),
\end{eqnarray}
We plug such ansatz into the HJ-equation and obtain the following series of the equations by the power expansion in $\lambda$. 
The ansatz is comprised of the series of the fields $f_k$ with arbitrary coefficients, $D^{(n)}$. The HJ-equation is an identical equation for the fields.
The equations are given by
\begin{eqnarray}
\label{floweq1}
\partial_\epsilon \Lambda(\epsilon)&=&-\frac{1}{2} \int d^5 k J_k(\epsilon)J_{-k}(\epsilon), \\
\label{floweq2}
\partial_\epsilon J_k(\epsilon)&=&-2J_k(\epsilon)D^{(2)}_{k,-k}(\epsilon), \\ \nonumber
\\ \nonumber
\label{floweq3}
\partial_\epsilon D^{(2)}_{(p,-p)}(\epsilon)&=&-\frac{1}{2}(4D^{(2)}_{(p,-p)}(\epsilon)D^{(2)}_{(-p,p)}(\epsilon)-p^2)-3\lambda\int d^5k J_{-k}(\epsilon)D^{(3)}_{(p,-p+k,-k)}(\epsilon) \\ 
\label{floweq4}
\partial_\epsilon D^{(3)}_{(k_1,k_2,k_3)}(\epsilon)&=&\frac{1}{4(2\pi)^{5/2}}-2\left( \sum_{j=1}^3 D^{(2)}_{k_j,-k_j}\right)(\epsilon) D^{(3)}_{k_1,k_2,k_3}(\epsilon)
\\ \nonumber
&-&4\lambda\int d^5k J_{-k}(\epsilon)D^{(4)}_{(k_1,k_2,-k_1-k_2+k,-k)}(\epsilon)  \\ 
\label{floweqn}
\partial_\epsilon D^{(n)}_{(k_1,...,k_{n})}(\epsilon)&=&
-2\left(\sum_{j=1}^nD^{(2)}_{(k_j,-k_j)}\right)(\epsilon)D^{(n)}_{(k_1,...,,k_{n-1},-\sum_{j=1}^{n-1}k_j)}(\epsilon)\\ \nonumber
-\frac{1}{2}\sum_{n^\prime=1}^{n-3}(n^\prime+2)(n-n^\prime)&{\mathcal P}&\left \{
D^{(n^\prime+2)}_{(k_1,...,k_{n^\prime+1},-\sum_{j=1}^{n^\prime+1}k_j)}(\epsilon)
D^{(n-n^\prime)}_{(k_{n^\prime+2},...,k_{n-1},-\sum_{j=1}^{n-1}k_j,\sum_{j=1}^{n^\prime+1}k_j)}(\epsilon)\right\}, \\ \nonumber
&-&\lambda(n+1)\int d^5k J_{-k}(\epsilon)D^{(n+1)}_{(k_1,...,k_{n-1},k-\sum_{j=1}^{n-1}k_j,-k)}
\\
\nonumber
{\ \ \rm \ for\ \ }n\geq 4,
\end{eqnarray}

\paragraph{$J_k=0$ solution:}
The simplest solutions are able to obtained when $J_k=0$ identically, which means that there is no single trace deformation at all. In such case, the boundary cosmological constant, $\Lambda=0$. 
The other parts of the solutions are given by
\begin{eqnarray}
\label{d-trace-sol}
D^{(2)}_{p,-p}(\epsilon)&=&\frac{1}{2}\frac{\partial_\epsilon f_p(\epsilon)}{f_p(\epsilon)}, \\
\label{t-trace-sol}
D^{(3)}_{(k_1,k_2,k_3	)}(\epsilon)&=&\frac{1}{4(2\pi)^{5/2}} \frac{\int^\epsilon \left( f_{k_1}(\epsilon^\prime)f_{k_2}(\epsilon^\prime)f_{k_3}(\epsilon^\prime)\right) d\epsilon^\prime+C^{(3)}_{k_1,k_2,k_3}}{f_{k_1}(\epsilon)f_{k_2}(\epsilon)f_{k_3}(\epsilon)}, \\
\label{n-trace-sol}
D^{(n)}_{(k_1,...,k_{n-1},-\sum_{j=1}^{n-1}k_j)}(\epsilon)&=&\frac{C^{(n)}}{\prod_{i=1}^n f_{k_i}(\epsilon)}-\frac{1}{2}\int ^\epsilon d\epsilon^\prime \left(\frac{\prod_{j=1}^n f_{k_j}(\epsilon^\prime)}{\prod_{l=1}^n f_{k_l}(\epsilon)}\right) \\ \nonumber
\times\sum_{n^\prime=1}^{n-3}(n^\prime+2)(n-n^\prime)&{\mathcal P}&\left\{
D^{(n^\prime+2)}_{(k_1,...,k_{n^\prime+1},-\sum_{j=1}^{n^\prime+1}k_j)}(\epsilon)
D^{(n-n^\prime)}_{(k_{n^\prime+2},...,k_{n-1},-\sum_{j=1}^{n-1}k_j,\sum_{j=1}^{n^\prime+1}k_j)}(\epsilon)\right\},
\end{eqnarray}
where
\begin{equation}
f_p(\epsilon)=C_p \cosh(|p|\epsilon+\theta_p), 
f_p(\epsilon)=C_p \sinh (|p|\epsilon+\theta_p), {\rm \ \ or \ \ }f_p(\epsilon)=C_p e^{\pm|p|\epsilon}
\end{equation}
where $C_p$ and $\theta_p$ are arbitrary, momentum $p$ dependent constants. 

The explicit form of the solution of the double trace operator $D^{(2)}_{p,-p}$ is given by
\begin{eqnarray}
D^{(2)}_{p,-p}(\epsilon)=\frac{|p|}{2}\tanh(|p|\epsilon+\theta_p), {\rm \ \ where\ \ } |D^{(2)}_{p,-p}(\epsilon)|\leq\frac{|p|}{2}, \\ \nonumber
D^{(2)}_{p,-p}(\epsilon)=\frac{|p|}{2}\coth(|p|\epsilon+\theta_p), {\rm \ \ where\ \ } |D^{(2)}_{p,-p}(\epsilon)|\geq\frac{|p|}{2}, \\ \nonumber
{\rm \ \ or \ \ }D^{(2)}_{p,-p}(\epsilon)=\pm\frac{|p|}{2}
\end{eqnarray}
It shows a trivial $IR$ fixed point when $\epsilon \rightarrow \infty$ and its $IR$ value 
is $D^{(2)}_{p,-p}(\infty)=\frac{|p|}{2}$. We note that the double trace operator is the square of the norm of the single trace operator, $O$. Therefore, if the double trace deformation is negative, it may spoil the unitarity of the deformed theory.

The multi-trace operators present diverse behaviors. Let us check the triple trace operator. Its explicit solution is given by
\begin{eqnarray}
D^{(3)}_{k_1,k_2,k_3}(\epsilon)&=&\frac{1}{4(2\pi)^{5/2}\prod_{i=1}^3\cosh(|k_i|\epsilon+\theta_i)}\left( \frac{C_p^{(3)}}{\prod_{i=1}^3C_{k_i}}
+\frac{1}{4}\frac{\sinh\left(\sum_{j=1}^3(|k_j|\epsilon+\theta_{k_j})\right)}{\sum_{l=1}^3|k_l|} \right. \nonumber \\ \nonumber
 &+&\left.\frac{1}{4}\sum_{j=1}^3\frac{\sinh\left(\sum_{l=1}^3(|k_l|\epsilon+\theta_{k_l})-2(|k_j|\epsilon+\theta_{k_j})\right)}{\sum_{m=1}^3|k_m|-2|k_j|} \right)  
\\ \nonumber
D^{(3)}_{k_1,k_2,k_3}(\epsilon)&=&\frac{1}{4(2\pi)^{5/2}\prod_{i=1}^3\sinh(|k_i|\epsilon+\theta_i)}\left( \frac{C_p^{(3)}}{\prod_{i=1}^3C_{k_i}}
+\frac{1}{4}\frac{\cosh\left(\sum_{j=1}^3(|k_j|\epsilon+\theta_{k_j})\right)}{\sum_{l=1}^3|k_l|} \right. \nonumber \\ \nonumber
 &-&\left.\frac{1}{4}\sum_{j=1}^3\frac{\cosh\left(\sum_{l=1}^3(|k_l|\epsilon+\theta_{k_l})-2(|k_j|\epsilon+\theta_{k_j})\right)}{\sum_{m=1}^3|k_m|-2|k_j|} \right),
\end{eqnarray}
The both of the solutions show their fixed point as
\begin{equation}
D^{(3)}_{k_1,k_2,k_3}(\infty)=\frac{1}{4(2\pi)^{5/2}(\sum_{i=1}^3|k_i|)}
\end{equation}
The multi trace operator solutions can be given by solving the equations that we addressed previously and its analytic forms are given in appendix.

\paragraph{Fixed points of multi trace operators in $J_k=0$ case:}
The fixed points of the multi trace operators can be obtained by requesting the equations(\ref{floweq3}),(\ref{floweq4}) and (\ref{floweqn}) at the fixed points. When the equation(\ref{floweq3}) is examined to find a fixed points under a condition that its left hand side vanishes i.e. $\partial_{\epsilon^\star} D^{(2)}_{p,-p}(\epsilon^\star)=0$, then the fixed points are given by
\begin{equation}
D^{(2)}_{p,-p}(\epsilon^\star)=\pm \frac{|p|}{2}
\end{equation}
By looking at the given solution of the double trace operator(\ref{d-trace-sol}), 
$D^{(2)}_{p,-p}(\epsilon^\star)= \frac{|p|}{2}$ is a fixed point at $\epsilon^\star=\infty$(so called IR fixed point) and 
$D^{(2)}_{p,-p}(\epsilon^\star)= -\frac{|p|}{2}$ is a fixed point at $\epsilon^\star=-\infty$(a fixed point in relatively UV region).

The equation(\ref{t-trace-sol}) shows fixed points of the triple trace operator as
\begin{equation}
D^{(3)}_{k_1,k_2,k_3}(\epsilon^\star_\pm)=\pm\frac{1}{4(2\pi)^{5/2}(\sum_{i=1}^3|k_i|)},
\end{equation}
where $\epsilon_\pm^\star=\pm\infty$. In general the fixed points of the $n$-trace operators are given by
the following condition
\begin{eqnarray}
\nonumber
D^{(n)}_{(k_1,...,k_{n})}(\epsilon^\star_\pm)&=&\mp\frac{1}{2(\sum_{i=1}^{n}|k_i|)}
\sum^{n-3}_{n^\prime=1}(n^\prime+2)(n-n^\prime){\mathcal P}\left\{D^{(n^\prime+2)}_{(k_1,...,k_{n^\prime+1},-\sum_{j=1}^{n^\prime+1}k_j)}(\epsilon^\star_\pm)\right.\\ 
&\times& \left.D^{(n-n^\prime)}_{(k_{n^\prime+2},...,k_{n-1},-\sum_{j=1}^{n-1}k_j,\sum_{j=1}^{n+1}k_j)}(\epsilon^\star_\pm)\right\}
\end{eqnarray}  
By utilizing such relation, for example,  one can derive the fixed points of the quadraple-trace operator as
\begin{eqnarray}
\nonumber
D^{(4)}_{k_1,k_2,k_3,k_4}(\epsilon^\star_\pm)&=&\mp\frac{3}{2^5(2\pi)^5(\sum_{i=1}^4|k_i|)}\left(\frac{1}{(|k_1|+|k_2|+|k_1+k_2|)(|k_3|+|k_4|+|k_3+k_4|)}\right. \\ 
&+&\left.({k_1 \leftrightarrow k_3})+({k_1 \leftrightarrow k_4})\right)
\end{eqnarray}



\paragraph{$J_k\neq0$ solutions}
In the case that we turn on the single trace operator, $J_k\neq0$, it is not easy to get their solutions since the equations become integral equations as well as the radial evolution of the $n$-ple trace operator depends on the $n+1$-ple trace operator through the terms $\sim \int d^5 kJ_k D^{(n+1)}_k$ in the equations. 
One way to solve such equations is by an assumption that the single trace term $J_k$ is parameterically small. We introduce a dimensionless small parameter $\alpha$, and assume that the single trace term is suppressed by the parameter.

The solutions of the single trace and the boundary cosmological constant are given by
\begin{equation}
J_k=\frac{\alpha B_k}{f_k(\epsilon)}, {\rm \ \ and \ \ }
\Lambda(\epsilon)=-\alpha^2\frac{1}{2}\int d^5k\frac{B_k B_{-k}}{f_k(\epsilon)f_{-k}(\epsilon)}, 
\end{equation}
where $B_k$ is arbitary boundary momentum $k_i$ dependent function. To get the $n$-ple trace operators, we try
the following form of ansatz,
\begin{equation}
\label{idontknow}
\mathbb D^{(n)}_{k_1,...,k_{n-1},-\sum^{n-1}_{j=1}k_j}(\epsilon)\equiv D^{(n)}_{k_1,...,k_{n-1},-\sum^{n-1}_{j=1}k_j}(\epsilon)+\alpha C^{(n)}_{k_1,...,k_{n-1},-\sum^{n-1}_{j=1}k_j}(\epsilon)+O(\alpha^2)
\end{equation}
where $\mathbb D^{(n)}(\epsilon)$ is the full solution which is a form of power expansion order by order in small parameter $\alpha$. $D^{(n)}(\epsilon)$ is the $n$-ple trace operator that we obtained in the previous section and $C^{(n)}(\epsilon)$ is the first order solution in the $\alpha$ respectively. 

It is very unlikely to get analytic solutions of $C^{(n)}(\epsilon)$. However, one certain statement is that the fixed points of the $n$-ple trace operator does not change by the first order solutions in $\alpha$ at $r=\epsilon^\star$, as long as the zeroth order solution 
of the $n+1$-ple trace operator's fixed points are finite at the given radial velue.

\section{Holographic 4-point function and conformal Ward identities}
\label{Holographic 4-point function and conformal Ward identities}

Conformal 2- and 3-point correlation functions are determined due to the global conformal invariance. 
In the momentum space as well as in the position space, such correlators  are investigated.
In \cite{Bzowski:2013sza}, the authors compute the 2- and 3-point functions in momentum space by imploying Fourier transform
from them in position space. The 2-point function the they obtain is given by
\begin{equation}
\langle O(p_1)O(p_2) \rangle=(2\pi)^d \delta^{(d)}(p_1+p_2) \langle\langle O(p_1)O(p_2) \rangle\rangle,
\end{equation}
where
\begin{equation}
\langle\langle O(p_1)O(p_2) \rangle\rangle=\frac{C_2 \pi^{d/2}2^{d-2\Delta}\Gamma\left( \frac{d-2\delta}{2} \right)}{\Gamma(\Delta)}|p|^{2\Delta-d}., 
\end{equation}
The $C_2$ is the coefficient of the 2-point function in position space, $\Delta$ is the conformal dimension of the operator $O(x)$, $\Gamma$ is the gamma function and the 
$d$ is the dimensionality of the spacetime. 
In the previous researches devoting to conformally coupled scalar in AdS spacetime and its holographic duals\cite{Jatkar:2013uga,Oh:2014nfa}, 
the authors compute the conformal 2-point correlation functions for the cases that $2\Delta-d=1$.
The 2-point functions have 
the forms of 
\begin{equation}
\langle\langle O(p_1)O(p_2) \rangle\rangle = C|p|,
\end{equation}
upto a coefficient $C$.

The form of the 3-point function is an integral form being given by
\begin{equation}
\langle\langle O(p_1)O(p_2)O(p_3) \rangle\rangle = C_3{p^{\Delta_1-\frac{d}{2}}_1}{p^{\Delta_2-\frac{d}{2}}_2}{p^{\Delta_3-\frac{d}{2}}_3}\int^\infty_0
dx{\ } x^{\frac{d}{2}-1}\left\{ \prod^3_{j=1} K_{\Delta_j-\frac{d}{2}}(p_j x) \right\},
\end{equation}
where the coefficient $C_3$ is 
\begin{equation}
C_3=\frac{C_{123}\pi^d 2^{4+\frac{3d}{4}-\Delta_t}}{\Gamma\left(\frac{\Delta_t-d}{2}\right)
\Gamma\left(\frac{\Delta_1+\Delta_2-\Delta_3}{2}\right)
\Gamma\left(\frac{\Delta_2+\Delta_3-\Delta_1}{2}\right)
\Gamma\left(\frac{\Delta_3+\Delta_1-\Delta_2}{2}\right)},
\end{equation}
and the $K_\nu$ represents the modified Bessel-K function.
The form of the 3-point function is not analytic form yet(except some special cases), which is so called``triple-K integral"\cite{Bzowski:2013sza}. 

In \cite{Jatkar:2013uga,Oh:2014nfa} and this note, the authors suggest the forms of the 3-point functions from a holographic computations. They study conformally coupled scalar theory in AdS background again. It turns out the model enjoies a reasonable self-interaction for the cases that $d=5$ and $d=3$(which are the boundary spacetime dimensions of the AdS space)
\footnote{In other cases, the self-interaction is not an integral power of the fields.}
. In such cases, the conformal dimensions corresponding of the exitations in tis AdS space are
$\Delta=3$ and $\Delta=2$ respectively.  

For the case that $d=5$, the 3-point function from the model(they perform minimal substraction and add no other terms on the conformal boundary) is
\begin{equation}
D^{(3)}_{k_1,k_2,k_3}(\epsilon^\star_\pm)=\pm\frac{1}{4(2\pi)^{5/2}(\sum_{i=1}^3|k_i|)},
\end{equation}
as addressed in this note, where absolutely the sum of all the momenta needs to be conserved. One can show that when $\Delta_1=\Delta_2=\Delta_3=3$ and $d=5$, the triple-K integral becomes the same form with this upto the overall coefficient. 

In the case that $\Delta=2$ and $d=3$\cite{Oh:2014nfa}, they find that the holographic computation of the 3-point function vanishes(again their prescription is minimal subtraction and adding no other manipulations on the conformal boundary). The triple-K integral suggest that 
\begin{equation}
\langle\langle O(p_1)O(p_2)O(p_3) \rangle\rangle \sim \int^\infty_0 \frac{dx}{x}\exp\left\{{-\left(\sum_{i=1}^3|p_i|\right)x}\right\},
\end{equation}
which has no momentum dependence and so it is a number.

There days, there are some of attampts to compute 4-point correlation function in momentum space\cite{Bzowski:2019kwd}.

\paragraph{Conformal Ward identities}
Dilatation Ward identity is $\mathcal D\left( p_j,\frac{\partial}{\partial p_j} \right)\Phi(p_1,...,p_n)=0$, where $\Phi$ is the $n$-point correlation function in momentum space, the momenta are constrained to accept momentum conservation $\sum_{j=1}^n \vec p_j=0$. The differential operator $\mathcal D$ is given by
\begin{equation}
\mathcal D=p^a \frac{\partial}{\partial p_a}+\Delta^\prime,
\end{equation}
where the $\Delta^\prime=-\sum_{i=1}^n\Delta_i + (n-1)d$.

Special conformal Wand identity is $\mathcal K^k\left(p_j,\frac{\partial}{\partial p_j}\right) \Phi(p_1,...,p_{n})=0$,
where $\Phi$ is the $n$-point correlation function in momentum space, the momenta are constrained to accept momentum conservation $\sum_{j=1}^n \vec p_j=0$. In fact, the operator $\mathcal K^k$ is an differential operator with respect to $n-1$ independent momenta. More preceisely, $\mathcal K^k\equiv\sum_{j=1}^{n-1} \mathcal K_j^k$ and 
\begin{equation}
\mathcal K^k_{j}=2(\Delta_j-d)\frac{\partial}{\partial p^k_j} + p_j^k\frac{\partial^2}{\partial p^a_j \partial p_j^a} - 2p_j^a\frac{\partial}{\partial p^k_j \partial p_j^a},
\end{equation}
where the $\Delta_j$ is the conformal dimension of the j-th operator.

If one applies this operator to a product of a certain two arbitrary functions of momenta, then it must be by chain rule as
\begin{eqnarray}
\mathcal K^k[A(p_l)B(p_m)]=\left\{ \mathcal K^k A(p_l) \right\} B(p_m) + A(p_l)\left\{ \mathcal K^k B(p_m) \right\} \\ \nonumber
+\sum_{j=1}^{n-1}\left( 2 p^k_j \frac{\partial A}{\partial p^a_j} \frac{\partial B}{\partial p^a_j}
-2 p^a_j \frac{\partial A}{\partial p^k_j} \frac{\partial B}{\partial p^a_j}
-2 p^a_j \frac{\partial A}{\partial p^a_j} \frac{\partial B}{\partial p^k_j}
\right)
\end{eqnarray}

\paragraph{4-point functions from conformally coupled scalar}
The holographic 4-point function is obtained from the conformally coupled scalar theory in AdS space\cite{Jatkar:2013uga,Oh:2014nfa}. For $d=3$ and $\Delta=2$ scalar operators, the 4-point function obtained from the model is
\begin{equation}
\langle O(p_1)O(p_2)O(p_3)O(p_4) \rangle=u^{-1}(4)\delta^{(4)}\left(\sum_{j=1}^n p_j\right),
\end{equation}
where 
\begin{equation}
u(n)=\sum_{j=1}^{n-1}|p_j|+\left|\sum_{j=1}^{n-1}p_j\right|.
\end{equation}
Namely, the $u(4)=|p_1|+|p_2|+|p_3|+|p_1+p_2+p_3|$. 

In fact, the object $u^{-1}(n)$ can be a candidate for $n$-point conformal scalar correlator for specific conformal dimensions. When one applies the special conformal Wand identity on the $u^{-1}(n)$, one gets in the end,
\begin{equation}
\label{u(n)-SCWI}
\mathcal K^k u^{-1}(n)=\frac{d-2\Delta+1}{u^2(n)}\left(\sum_{j=1}^{n-1}\frac{p_j^k}{|p_j|}\right)
+\frac{2(d-\Delta)(n-1)-d-3}{u^2(n)}
\left(\frac{\sum_{j=1}^{n-1} p_j^k}{\sum_{l=1}^{n-1}|p_l|}\right)
\end{equation} 
For the $u^{-1}(n)$ to become a conformal correlator, the right hand side of the (\ref{u(n)-SCWI}) should vanish. Then,  we have two algebraic equations. After a little computations, the equations become
\begin{equation}
d=2\Delta-1,{\ \ \ \rm and \ \ } \Delta=\frac{2}{n-2}+1.
\end{equation}
Together with this, one needs to consider the dilatation Ward identity too. This gives a more algebraic constraint, which is
\begin{equation}
\Delta=\frac{(n-1)d-1}{n}.
\end{equation}
The possible and reasonable(both the spacetime dimension, $d$ and the $n$ are integral) solutions are listed below:
\begin{itemize}
\item $d=2$, $\Delta=\frac{3}{2}$ and $n=6$, so the $u^{-1}(6)$ is a conformal 6-point function for the scalar operators with conformal dimension $\Delta=\frac{3}{2}$ in $d=2$.
\item $d=3$, $\Delta=2$ and $n=4$, so the $u^{-1}(4)$ is a conformal 4-point function for the scalar operators with conformal dimension $\Delta=2$ in $d=3$. This is what we get in holographic 4-point function from the conformally coupled scalar model.
\item $d=5$, $\Delta=3$ and $n=3$, so the $u^{-1}(3)$ is a conformal 3-point function for the scalar operators with conformal dimension $\Delta=3$ in $d=5$. This is what we get in holographic 3-point function from the conformally coupled scalar model in this note.
\end{itemize}

The 4-point function that we get in this note for the $5$-dimensional conformal field theory( obtained from the conformally coupled scalar model in AdS$_6$) is given by
%
\begin{eqnarray}
\label{4-finction-point}
D^{(4)}_{k_1,k_2,k_3,k_4}(\epsilon^\star_\pm)&=&\mp\frac{3}{2^5(2\pi)^5u(4)u(3)v(4,1)}+({k_1 \leftrightarrow k_3})+({k_2 \leftrightarrow k_3}),
\end{eqnarray}
where
\begin{equation}
v(E,n)=\left| \sum_{i=1}^{E-1-n}P_i \right|+\sum_{j=E-n}^{E-1}|p_j|+\left| \sum_{k=1}^{E-1}p_k \right|.
\end{equation}
The first term in (\ref{4-finction-point}) is s-channel, the second is t- and the last term is u- channel in order.

When one applies the special conformal Ward identity on the first term of the holographic 4-point function, which gives
\begin{equation}
\label{SCWI-result}
\mathcal K^k D^{(4)}_{k_1,k_2,k_3,k_4} = 2\frac{\{u(3)-u(4)\}\{u(4)+v(4,1)\}}{u^2(3)u^2(4)v^2(4,1)}\left\{ \frac{(p_1+p_2)^k}{|p_1+p_2|} +  \frac{(p_1+p_2+p_3)^k}{|p_1+p_2+p_3|}\right\}
\end{equation}
\paragraph{Discussion}
The result above that the (s-channel)4-point function is not a conformal correlation function since it does not satisfy the special confromal Ward identity in general. However, in a certain limit, it does. The right hand side of (\ref{SCWI-result}) is proportional to the factor, $u(3)-u(4)=|p_1+p_2|-|p_3|-|p_1+p_2+p_3|$. Considering momentum conservation, this becomes $|p_3+p_4|-|p_3|-|p_4|$, which vanishes when $\vec p_3$ and $\vec p_4$ are colinear. There is another factor $\left\{ \frac{(p_1+p_2)^k}{|p_1+p_2|} +  \frac{(p_1+p_2+p_3)^k}{|p_1+p_2+p_3|}\right\}=\hat n_{12}-\hat n_4$, where $\hat n_{12}$ is a unit vector alnog $\vec p_1+\vec p_2$ and $\hat n_4$ is a nunit vector along $\vec p_4$. If a condition 
$\hat n_{12}=\hat n_4$, then the right hand side of (\ref{SCWI-result}) vanishes. This means that
$\vec p_1+\vec p_2$ and $\vec p_4$ are colinear.

If one sum up all posible channels, then the above argument is not hold. The only possible limit which makes
the 4-point function conformal, is that $\vec p_i$ for $i=1,2,3$ and $4$ are colinear. In this limit, the 4-point function effectively becomes
\begin{equation}
\label{u3}
\sim\frac{1}{u^3(4)},
\end{equation}
which is expected in \cite{Oh:2020izq}. In \cite{Oh:2020izq}, it is proved that $n$-point correlation function among the scalar operators $O_{\Delta}$, where $\Delta=\frac{d+1}{2}$ is given by
\begin{equation}
\Phi(p_1,p_2,...,p_n)=\frac{1}{\left(
\sum_{i=1}^{n-1}|p_i|+\left| \sum_{j=1}^{n-1} p_j \right|
\right)^{(n-1)d-\frac{n}{2}(d+1)}},
\end{equation}
and the (\ref{u3}) is recovered when $n=4$ and $d=5$.
Again, $d$ is spatial dimensionality that the theory defined on and the space is Eucliean.

The colinear limit of the 4-point function may be related to the colinear divergences of the solution of the equation of motion of the conformally coupled scalar theory in AdS space. To remove the colinear divergences of the soluton, we add the homogeneous solutions of the equation of motion to regulariz it. In the prevous computation, we choose such a scheme. In such scheme, it is very clear that the holographic correlation function from the conformally coupled scalar theory is free from the colinear divergneces. However, we point out the forms of the boundary correlation functions are scheme-independent.

We also point out that one may recognize that in the colinear limit, the $n$-point holographic correlation functions that we get, or the multi-trace deformations in the fixed points $\epsilon=\epsilon^\star_+$ of the scalar operators $O$ effectively become the conformal correlation functions given in \cite{Oh:2020izq}. 
More precisely, 
\begin{equation}
\langle O({p_1})O({p_2})..O({p_n})\rangle=\mathcal D^{(n)}_{p_1,p_2,...,p_n}(\epsilon^\star_+)\sim\frac{C_n}{\left(   \sum_{i=1}^{n-1}|p_i|+|\sum_{i=1}^{n-1}p_i| \right)^{2n-5}},
\end{equation}
where the colinear limit denotes that all the external momenta $p_i$s for $i=1,...,n$ are alined in the same direction and so it is satisfied that $\sum_{i=1}^{n}|p_i|=|\sum_{i=1}^{n}p_i|$. Therefore, in some sense, one may say that the conformally coupled scalar theory in $AdS_6$ produces conformal correlation functions of a scalar operator $O$ with $\Delta=3$ in 5-dimensinal Euclidean space as its dual conformal field theory.

We also expect that the conformally coupled scalar theory in $AdS_4$ will produce $n$-point conformal correlation functions as a well-posed dual gravity theory of a certain $3$-dimensional CFT deformed with a scalar operator $O$ with its conformal dimension is $\Delta=2$ in a colinear limit. It could be an excercise for my students.

\section*{Acknowledgement}
J.H.O would like to thank his $\mathcal W.J.$ and $\mathcal Y.J.$ This work was supported by the National Research Foundation of Korea(NRF) grant funded by the Korea government(MSIP) (No.2016R1C1B1010107). This work is also partially supported by Research Institute for Natural Sciences, Hanyang University.

\section{Appendices}
\subsection{Appendix A: Application of special conformal Ward identity to the holographic 4-point function}

\begin{eqnarray}
\mathcal K^k\left(\frac{1}{u(4)u(3)v(4,1)} \right)=\frac{1}{u(3)v(4,1)}\mathcal K^k\left( 
\frac{1}{u(4)} \right)+\frac{1}{u(3)u(4)}\mathcal K^k\left( 
\frac{1}{v(4,1)} \right) \\ \nonumber
+ {\ \ \rm A-type\ terms\ \ } + {\ \ \rm B-type\ terms\ \ } + {\ \ \rm C-type\ terms\ \ },
\end{eqnarray}
where
\begin{eqnarray}
&{\ }&{\ \ \rm A-type\ terms\ \ }= 2\sum_{l=1}^3p_l^k \frac{1}{u(3)} \frac{\partial}{\partial p_l^a }\left( \frac{1}{u(4)} \right)\frac{\partial}{\partial p_l^a }\left( \frac{1}{v(4,1)} \right) \\ \nonumber
&-&2\sum_{l=1}^3p_l^a \frac{1}{u(3)} \frac{\partial}{\partial p_l^k }\left( \frac{1}{u(4)} \right)\frac{\partial}{\partial p_l^a }\left( \frac{1}{v(4,1)} \right)
-2\sum_{l=1}^3p_l^a \frac{1}{u(3)} \frac{\partial}{\partial p_l^a }\left( \frac{1}{u(4)} \right)\frac{\partial}{\partial p_l^k }\left( \frac{1}{v(4,1)} \right),
\end{eqnarray}
\begin{eqnarray}
&{\ }&{\ \ \rm B-type\ terms\ \ }= 2\sum_{l=1}^3p_l^k \frac{1}{u(4)} \frac{\partial}{\partial p_l^a }\left( \frac{1}{u(3)} \right)\frac{\partial}{\partial p_l^a }\left( \frac{1}{v(4,1)} \right) \\ \nonumber
&-&2\sum_{l=1}^3p_l^a \frac{1}{u(4)} \frac{\partial}{\partial p_l^k }\left( \frac{1}{u(3)} \right)\frac{\partial}{\partial p_l^a }\left( \frac{1}{v(4,1)} \right)
-2\sum_{l=1}^3p_l^a \frac{1}{u(4)} \frac{\partial}{\partial p_l^a }\left( \frac{1}{u(3)} \right)\frac{\partial}{\partial p_l^k }\left( \frac{1}{v(4,1)} \right),
\end{eqnarray}
and
\begin{eqnarray}
&{\ }&{\ \ \rm C-type\ terms\ \ }= 2\sum_{l=1}^3p_l^k \frac{1}{v(4,1)} \frac{\partial}{\partial p_l^a }\left( \frac{1}{u(3)} \right)\frac{\partial}{\partial p_l^a }\left( \frac{1}{u(4)} \right) \\ \nonumber
&-&2\sum_{l=1}^3p_l^a \frac{1}{v(4,1)} \frac{\partial}{\partial p_l^k }\left( \frac{1}{u(3)} \right)\frac{\partial}{\partial p_l^a }\left( \frac{1}{u(4)} \right)
-2\sum_{l=1}^3p_l^a \frac{1}{v(4,1)} \frac{\partial}{\partial p_l^a }\left( \frac{1}{u(3)} \right)\frac{\partial}{\partial p_l^k }\left( \frac{1}{u(4)} \right).
\end{eqnarray}


Each term is given by
\begin{equation}
\mathcal K^k\left( 
\frac{1}{u(4)} \right)=\frac{4}{u^2(4)} \frac{(p_1+p_2+p_3)^k}{|p_1+p_2+p_3|},
\end{equation}
which is given in (\ref{u(n)-SCWI}).

\begin{equation}
\mathcal K^k\left( 
\frac{1}{v(4,1)} \right)=\frac{4}{v^2(4,1)}\left\{ \frac{(p_1+p_2)^k}{|p_1+p_2|} +  \frac{(p_1+p_2+p_3)^k}{|p_1+p_2+p_3|}\right\}
\end{equation}

\begin{equation}
{\ \ \rm A-type\ terms\ \ }=\frac{2}{u(3)u^2(4)v^2(4,1)}\left\{  \frac{(p_1+p_2)^k}{|p_1+p_2|}(v(4,1)-u(4)) -  \frac{(p_1+p_2+p_3)^k}{|p_1+p_2+p_3|}(v(4,1)+u(4))   \right\}
\end{equation}

\begin{equation}
{\ \ \rm B-type\ terms\ \ }=-\frac{2}{u^2(3)v^2(4,1)}\left(  \frac{(p_1+p_2)^k}{|p_1+p_2|} +  \frac{(p_1+p_2+p_3)^k}{|p_1+p_2+p_3|}   \right)
\end{equation}

\begin{equation}
{\ \ \rm C-type\ terms\ \ }=-\frac{2}{u^2(3)u(4)v(4,1)}\left(  \frac{(p_1+p_2)^k}{|p_1+p_2|} +  \frac{(p_1+p_2+p_3)^k}{|p_1+p_2+p_3|}   \right)
\end{equation}

\end{document}